\documentclass[english,superscriptaddress,showpacs,showkeys]{revtex4}
\usepackage{amsmath}
\usepackage{graphicx}
\usepackage{amssymb}
\usepackage{esint}

\makeatletter

\providecommand{\tabularnewline}{\\}

\@ifundefined{textcolor}{}
{%
 \definecolor{BLACK}{gray}{0}
 \definecolor{WHITE}{gray}{1}
 \definecolor{RED}{rgb}{1,0,0}
 \definecolor{GREEN}{rgb}{0,1,0}
 \definecolor{BLUE}{rgb}{0,0,1}
 \definecolor{CYAN}{cmyk}{1,0,0,0}
 \definecolor{MAGENTA}{cmyk}{0,1,0,0}
 \definecolor{YELLOW}{cmyk}{0,0,1,0}
 }

\makeatletter

\providecommand{\tabularnewline}{\\}

\makeatletter

\newcommand{\ba}{\begin{array}{l}}\newcommand{\ea}{\end{array}}\renewcommand{\citet}[1]{\cite{#1}}\@ifundefined{definecolor}{\@ifundefined{definecolor}{\@ifundefined{definecolor}
 {\usepackage{color}}{}
}{}}{}

\makeatletter

\usepackage[hyperref]{hyperref}
\hypersetup{pdfauthor={K. Goeke, M. Musakhanov, M. Siddikov}, pdftitle={QCD isospin breaking ChPT low-energy constants from the instanton
vacuum}, pdfsubject={Chiral lagrangian}, pdfkeywords={Instanton vacuum, Large-Nc expansion, Chiral symmetry, Chiral lagrangian,
Pion properties}}

\makeatother

\renewcommand{\[}{\begin{equation}}
\renewcommand{\]}{\end{equation}}

\makeatother

\makeatother

\usepackage{babel}

\begin{document}

\title{QCD isospin breaking ChPT low-energy constants from the instanton
vacuum}

\author{K. Goeke}

\email{Klaus.Goeke@tp2.rub.de}

\affiliation{Institut f\"{u}r Theoretische Physik II, Ruhr-Universit\"{a}t-Bochum, D-44780
Bochum, Germany}

\author{M. Musakhanov}

\email{Yousuf@uzsci.net}

\affiliation{Theoretical Physics Department, National University of Uzbekistan,
Tashkent 100174, Uzbekistan}

\author{M. Siddikov}

\email{Marat.Siddikov@usm.cl}

\affiliation{Departamento de F\'{i}sica, Centro de Estudios Subat\'{o}micos, y Centro
Cient\'{i}fico - Tecnol\'{o}gico de Valpara\'{i}so, Universidad T\'{e}cnica
Federico Santa Mar\'{i}a, Casilla 110-V, Valpara\'{i}so, Chile}

\preprint{USM-TH-257}
\begin{abstract}
In the framework of the instanton vacuum model we evaluate the Chiral
Perturbation Theory (ChPT) low-energy constants $h_{3},l_{7}$. We
found that in the instanton vacuum model the constant $l_{7}$ is
very sensitive to the shape of the instanton and the instanton vacuum
parameters. We evaluated the constant $l_{7}$ for two different zero-mode
profiles and as a function of the average instanton size $\rho$ and
inter-instanton distance $R$. Our result agrees with an old {}``order
of magnitude'' estimate of this constant from~\cite{Gasser:1983yg}.
The obtained value of $l_{7}$ implies that the pure QCD contribution
to the pion mass difference is small, $\sim1\%$ of the observed experimental
value.
\end{abstract}

\pacs{
11.10.Lm, 
11.15.Kc, 
11.15.Pg, 
11.30.Rd 
12.39.Fe 
}

\keywords{Instanton vacuum, Large-Nc expansion, Chiral symmetry, Chiral lagrangian,
Pion properties}

\maketitle

\section{Introduction}

The spontaneous breaking of chiral symmetry (S$\chi$SB) is one of
the most important phenomena of hadron physics. It defines the properties
of all the light mesons and baryons. Using the general idea of chiral
symmetry, it was proposed in~\cite{Gasser:1983yg} to use a phenomenological
lagrangian, which has a form of the infinite series in the pion momenta
$p^{2}$ and mass $M_{\pi}^{2}$. The low-energy constants of the
series expansion (LEC's) are the free parameters which encode the
low-energy physics in a model-independent way. Up to now they were
extracted phenomenologically from the experimental data, or from the
lattice calculations ((MILC, ETM, JLQCD, RBC/UKQCD, PACS-CS)\cite{Bowman:2004xi,Chu:vi,DeGrand:2001tm,Negele:1998ev}
within so-called Chiral Perturbation Theory (ChPT).

One of the low-energy constants $l_{7}$ is particularly interesting
since it encodes the {}``pure QCD'' part of the $SU(2)$ isospin
symmetry breaking (\emph{i.e.} part which is due to $u-$ and $d-$quark
current mass difference, $m_{u}-m_{d}$).~ For example, the QCD part
of the pion mass difference $m_{\pi^{+}}^{2}-m_{\pi^{0}}^{2}$ has
a form~\cite{Gasser:1983yg}
\begin{equation}
\left(m_{\pi^{+}}^{2}-m_{\pi^{0}}^{2}\right)_{QCD}=\frac{2B^{2}}{F^{2}}l_{7}\left(m_{u}-m_{d}\right)^{2},\label{eq:PiMassSplit_QCD}
\end{equation}
where $B$ and $F$ are the leading order parameters in the chiral
lagrangian, and $m_{u},m_{d}$ are the current quark masses. While
experimentally the isospin breaking effects are known to a very high
precision, separation of these effects on the {}``pure QCD'' and
electromagnetic parts has ambiguities and has been a subject of intensive
debates~\cite{Bijnens:1993ae,Bijnens:1996kk,Rathske,Rusetsky:2009ic,Gasser:1983yg}.
From phenomenology the constant $l_{7}$ is known only with an {}``order
of magnitude'' estimate~\cite{Gasser:1983yg},\begin{equation}
l_{7}\sim5\times10^{-3}.\label{eq:L7-phenom-value}\end{equation}
For this reason it makes sense to estimate this contribution in the
framework of a reliable model.

QCD instanton vacuum model, often refered to as the instanton liquid
model, provides a very natural nonperturbative explanation of the
S$\chi$SB~\cite{'tHooft:1973jz,'tHooft:1976fv,Diakonov:1983hh,Diakonov:1985eg,Diakonov:1995qy,Diakonov:2002fq,Kim:2004hd,Kim:2005jc,LangackerPagels73,Lee:sm,Musakhanov:1998wp,Musakhanov:2002xa}.
It provides a consistent framework for description of the pions and
thus may be used for evaluation of the low energy constants. Due to
instanton-induced nonlinear interaction all the quark and meson loop
integrals are regularized by the natural scale $\mu\sim\rho^{-1}\sim600$~MeV
in Pauli-Villars scheme~\cite{Diakonov:1985eg}, where $\rho$ is
the average size of the instanton. This means that all the scale-dependent
quantities, such as the quark condensate $\left\langle \bar{q}q\right\rangle \equiv\left\langle \bar{u}u\right\rangle +\left\langle \bar{d}d\right\rangle $
and the difference~$\delta\left\langle \bar{q}q\right\rangle \equiv\left\langle \bar{u}u\right\rangle -\left\langle \bar{d}d\right\rangle $,
are given at the scale~$\mu$. Remarkably, the constant $l_{7}$
does not depend on the scale $\mu$. Recently \cite{Goeke:2007bj}
it has been shown that this approach is able to give results consistent
with phenomenological and lattice estimates for the constants $\bar{l}_{3},\bar{l}_{4}$,
providing current quark mass dependencies of the pion mass $m_{\pi}$
and pion decay constant $F_{\pi}$.

In this paper we would like to apply the instanton vacuum model for
the evaluation of the constant $l_{7}$. 
 We extract the constant $l_{7}$ from the correlator $\langle P^{3}(x)P^{0}(0)\rangle$
using the relation~\cite{Gasser:1983yg} \begin{equation}
\int d^{4}x\, e^{iqx}\langle P^{3}(x)P^{0}(0)\rangle=\frac{G_{\pi}\tilde{G}_{\pi}}{m_{\pi}^{2}-q^{2}}+\mathcal{O}\left(q^{2}\right)=\frac{8B^{3}\left(m_{u}-m_{d}\right)}{q^{2}-m_{\pi}^{2}}l_{7}+\mathcal{O}\left(m,q^{2}\right).\label{eq:P3P0Definition-1}\end{equation}
 From the Eqn.~(\ref{eq:P3P0Definition-1}) we may see that evaluations
may be done in the limit $m\equiv\frac{m_{u}+m_{d}}{2}\to0,$ and
make only expansion over \[
\delta m\equiv\left(m_{u}-m_{d}\right).\]

The paper is organized as follows. In Section~\ref{sec:GapEqn} we
discuss the general framework used for evaluation and write out the
next-to-leading order (NLO) gap equation in the presence of the current
mass split $\delta m$, which are needed for evaluation of the dynamical
mass split $\delta M\equiv M_{u}-M_{d}.$ In Section~\ref{sec:Propagators}
we write out explicit expressions for the quark and meson propagators.
In Section~\ref{sec:QuarkCondensate} we evaluate the effects of
the mass split $\delta m$ on the quark condensate, $\delta\langle\bar{q}q\rangle=\langle\bar{q}q\rangle_{u}-\langle\bar{q}q\rangle_{d}$
and extract the constant $h_{3}$. In Section~\ref{sec:P3P0Correlator}
we evaluate the correlator $\langle P^{3}(x)P^{0}(0)\rangle$ and
extract the constant $l_{7}$. In Section~\ref{sec:Conclusion} we
discuss obtained results, their uncertainty limits and draw conclusions.

\section{Instanton vacuum model}

The instanton vacuum model is based on the assumptions that the QCD
vacuum may be considered as a dilute gas of instantons and antiinstantons,
and the number of colors $N_{c}$ is asymptotically large, $N_{c}\to\infty$
(see the reviews \cite{Schafer:1996wv,Diakonov:2002fq}). While in
general the sizes and local density of the instanton gas may be arbitrary,
inter-instanton interaction stabilize these parameters. As it has
been discussed in~\cite{Goeke:2007bj}, the $1/N_{c}$-suppressed
corrections due to the finite size distribution are indeed quite small,
even for $N_{c}=3$. Phenomenological, variational and lattice estimates
lead to average instanton size $\rho\sim0.3\, fm$ and inter-instanton
distance $R\sim1\, fm$~\cite{Diakonov:1983hh}.

The partition function in the field of external scalar and pseudoscalar
currents~$s=\left(s_{0}+\vec{s}\vec{\tau}\right)$ and $p=\left(p_{0}+\vec{p}\vec{\tau}\right)$
has a form~\cite{Goeke:2007bj}

\begin{eqnarray}
Z_{N}[s_{0},\sigma,s_{0},\vec{p},\vec{s},p_{0}] & = & \int d\lambda\exp\left(-\Gamma_{eff}[s_{0},\lambda,\sigma,s_{0},\vec{p},\vec{s},p_{0},\vec{\sigma}_{v},\eta_{v},\vec{u}]\right),\label{eq:Zma}\\
\Gamma_{eff} & = & S+\Gamma_{eff}^{mes},\label{eq:Z0}\\
S & = & \frac{N}{V}\,\ln\lambda+2\int d^{4}x\sum\Phi_{i}^{2}(x)-Tr\ln\left(\frac{\hat{p}+is_{0}+\vec{s}\cdot\vec{\tau}+p_{0}\gamma_{5}+i\vec{p}\cdot\vec{\tau}\gamma_{5}+i\, c\, F\Phi\cdot\Gamma F}{\hat{p}+is_{0}+\vec{s}\cdot\vec{\tau}+p_{0}\gamma_{5}+i\vec{p}\cdot\vec{\tau}\gamma_{5}}\right),\label{eq:Z1}\end{eqnarray}
The nonlocal formfactors $F(p)$ in the meson-quark interaction vertices
come from the instanton-induced nonlocal interactions. Together with
the factor $\left(\hat{p}+is_{0}+\vec{s}\cdot\vec{\tau}+p_{0}\gamma_{5}+i\vec{p}\cdot\vec{\tau}\gamma_{5}\right)$
in denominator, which subtracts the divergent high-frequency modes,
they guarantee finite results for all the observables in the instanton
vacuum model. As it was discussed in~\cite{Diakonov:1983hh,Diakonov:1985eg},
the divergent high-ferquency modes are responsible for renormalization
of the parameters of the model. In what follows, we will fix them
at the scale $\mu\sim\rho^{-1}\sim600$~MeV in the Pauli-Villars
scheme~\cite{Diakonov:1983hh,Diakonov:1985eg}.

The meson-loop correction~$\Gamma_{eff}^{mes}$ to the effective
action is given as \begin{align}
\Gamma_{eff}^{mes}[m,\lambda,\sigma] & =\frac{1}{2}Tr\ln\left(4\delta_{ij}+\frac{1}{\sigma^{2}}Tr\left(\frac{c\left(\lambda\right)F^{2}(p)}{\hat{p}+is_{0}+\vec{s}\cdot\vec{\tau}+p_{0}\gamma_{5}+i\vec{p}\cdot\vec{\tau}\gamma_{5}+i\, c\, F\Phi\cdot\Gamma F}\Gamma_{i}\times\right.\right.\label{eq:Z2}\\
 & \left.\left.\frac{c\left(\lambda\right)F^{2}(p)}{\hat{p}+is_{0}+\vec{s}\cdot\vec{\tau}+p_{0}\gamma_{5}+i\vec{p}\cdot\vec{\tau}\gamma_{5}+i\, c\, F\Phi\cdot\Gamma F}\Gamma_{j}\right)\right),\nonumber \\
\Phi\cdot\Gamma= & \left(\sigma+i\gamma_{5}\vec{\tau}\vec{\phi}+i\vec{\tau}\vec{\sigma}+\gamma_{5}\eta\right),\end{align}
 where $c(\lambda)=\frac{(2\pi\rho)^{2}\sqrt{\lambda}}{2g},$$g^{2}=\frac{(N_{c}^{2}-1)2N_{c}}{2N_{c}-1}$
is a color factor, $\Gamma=\{1,\gamma_{5},i\vec{\tau},i\vec{\tau}\gamma_{5}\}$
is a set of matrices corresponding to quantum numbers of mesons present
in the model, and we will use for the corresponding components of
the field $\Phi$ the notations $\Phi=\{\sigma,\eta,\vec{\sigma},\vec{\phi}\}$.
In contrast to NJL model, the variable $\lambda$ is a dynamical degree
of freedom but not the parameter of the lagrangian. The current masses
of the quarks come into play via constant external currents, viz.
\begin{eqnarray}
s_{0} & = & \frac{m_{u}+m_{d}}{2},\\
s_{3} & = & \frac{m_{u}-m_{d}}{2}.\end{eqnarray}
 Notice that with respect to chiral transformations, the mesons may
be separated onto two independent doublets $\left(\sigma,\vec{\phi}\right)$
and $\left(\eta,\vec{\sigma}\right).$ The first doublet~$\left(\sigma,\vec{\phi}\right)$
corresponds to the pion field $U=\left(u_{0},\vec{u}\right)$ in the
notations of~\cite{Gasser:1983yg}, and the second doublet $\left(\eta,\vec{\sigma}\right)$
is an additional degree of freedom which is absent in the chiral lagrangian.
Now we are going to demonstrate explicitly on the example of the constant
$l_{7}$ that this additional degree of freedom $\left(\eta,\vec{\sigma}\right)$
gives an essential contribution to the constant $l_{7}$ . As usual,
the external currents $\left(s_{0},\,\vec{s},\, p_{0},\,\vec{p}\right)$
generate nonzero vacuum averages of the fields $\left\langle \vec{\sigma}\right\rangle =\vec{\sigma}_{v},$
$\left\langle \eta\right\rangle =\eta_{v}$ and $\left\langle \left(\sigma,\vec{\phi}\right)\right\rangle =U=\left(u_{0},\vec{u}\right)$.

Due to the chiral symmetry expansion of the $\Gamma_{eff}$ yields
the general structure \begin{eqnarray}
 &  & \Gamma_{eff}[\lambda,\sigma,\vec{s},p_{0},\vec{\sigma}_{v},\eta_{v},u_{i}]=\Gamma_{eff}[m,\lambda,\sigma,\vec{s}=0,p_{0}=0,\vec{\sigma}_{v}=0,\eta_{v}=0,u_{i}=0]\label{structure}\\
 &  & +\mathcal{A}\left(\left(\partial u_{0}\right)^{2}+\left(\partial\vec{u}\right)^{2}\right)+\mathcal{B}\left(s_{0}u_{0}+\vec{p}\vec{u}\right)+\mathcal{C}\left(s_{0}p_{0}+\vec{p}\vec{s}\right)^{2}+\mathcal{D}\left(s_{0}\eta_{v}+\vec{p}\vec{\sigma}_{v}\right)^{2}+a\left(p_{0}^{2}+{\vec{s}}^{2}\right)+b\left(p_{0}\eta_{v}+\vec{s}\vec{\sigma}_{v}\right)\nonumber \\
 &  & +c\left(\eta_{v}^{2}+{\vec{\sigma}_{v}}^{2}\right)+d\left(u_{0}p_{0}+\vec{u}\vec{s}\right)^{2}+e\left(u_{0}p_{0}+\vec{u}\vec{s}\right)\left(u_{0}\eta_{v}+\vec{u}\vec{\sigma}_{v}\right)+f\left(u_{0}\eta_{v}+\vec{u}\vec{\sigma}_{v}\right)^{2}+\mathcal{O}\left(s^{6},p^{6}\right),\nonumber \end{eqnarray}
 where we omitted the terms containing derivatives of the fields,
since the external currents are constants, the constants $\mathcal{A}-\mathcal{D},\, a-f$
should be evaluated with account of NLO corrections. The vacuum equations
which follow from (\ref{structure}) are \begin{eqnarray}
 &  & \frac{\partial\Gamma_{eff}[m,\lambda,\sigma_{v},\vec{s}=0,p_{0}=0,\vec{\sigma}_{v}=0,\eta_{v}=0]}{\partial\lambda}=\frac{\partial\Gamma_{eff}[m,\lambda,\sigma_{v},\vec{s}=0,p_{0}=0,\vec{\sigma}_{v}=0,\eta_{v}=0]}{\partial\sigma_{v}}=0\label{eq:SigmaLambdaGap}\\
 &  & \frac{\partial\Gamma_{eff}[m,\lambda,\sigma,\vec{s},p_{0},\vec{\sigma}_{v},\eta_{v},u_{i}]}{\partial\sigma_{v,i}}=\frac{\partial\Gamma_{eff}[m,\lambda,\sigma,\vec{s},p_{0},\vec{\sigma}_{v},\eta_{v},u_{i}]}{\partial\eta_{v}}=\frac{\partial\Gamma_{eff}[m,\lambda,\sigma,\vec{s},p_{0},\vec{\sigma}_{v},\eta_{v},u_{i}]}{\partial u_{i}}=0\label{eq:SigmaEtaGap}\end{eqnarray}
 The coefficients $\mathcal{A},\mathcal{B}$ are relevant for the
2-point correlators with intermediate pions and $\mathcal{A}\sim F^{2}$
and $\mathcal{B}\propto M_{\pi}^{2}$ in $m_{u}=m_{d}$ limit. The
constants $\mathcal{B},\mathcal{C},\mathcal{D}$ are irrelevant to
our problem since they are constants in front of the term with chiral
doublet $\chi=\left(s_{0},\vec{p}\right)$ which we put to zero in
the current paper.

The Eqns~(\ref{eq:SigmaLambdaGap}) are responsible for the dynamical
mass generation and will be discussed in the next section. The Eqns~(\ref{eq:SigmaEtaGap})
may be explicitly written as \begin{eqnarray}
 &  & \frac{\partial\Gamma_{eff}[m,\lambda,\sigma,\vec{s},p_{0},\vec{\sigma}_{v},\eta_{v},u_{i}]}{\partial\eta_{v}}=bp_{0}+2c\eta_{v}+eu_{0}\left(u_{0}p_{0}+\vec{u}\vec{s}\right)+2fu_{0}\left(u_{0}\eta_{v}+\vec{u}\vec{\sigma}_{v}\right)=0\label{eq:GapEtaExplicit}\\
 &  & \frac{\partial\Gamma_{eff}[m,\lambda,\sigma,\vec{s},p_{0},\vec{\sigma}_{v},\eta_{v},u_{i}]}{\partial\sigma_{v,i}}=b\vec{s}_{v}+2c\vec{\sigma}_{v}+e\vec{u}\left(u_{0}p_{0}+\vec{u}\vec{s}\right)+2f\vec{u}\left(u_{0}\eta_{v}+\vec{u}\vec{\sigma}_{v}\right)=0\label{eq:GapSigmaExplicit}\end{eqnarray}
 Multiplying Eqn~(\ref{eq:GapEtaExplicit}) on $u_{0}$ and Eqn.~(\ref{eq:GapSigmaExplicit})
on $\vec{u}$ and adding results, we may find: \begin{eqnarray}
u_{0}\eta_{v}+\vec{u}\vec{\sigma}_{v} & = & -\frac{b+e}{2\left(c+f\right)}\left(u_{0}p_{0}+\vec{u}\vec{s}\right).\label{eq:GapRes1}\end{eqnarray}
 Repeating the same trick with $p_{0}$ and $\vec{s}$, we may get
\begin{eqnarray}
p_{0}\eta_{v}+\vec{s}\vec{\sigma}_{v} & = & -\frac{1}{2c}\left[b\left(p_{0}^{2}+{\vec{s}}^{2}\right)+\left(e-f\frac{b+e}{c+f}\right)\left(u_{0}p_{0}+\vec{u}\vec{s}\right)^{2}\right],\label{eq:GapRes2}\end{eqnarray}
 and repeating the same trick with $\eta_{v}$ and $\vec{\sigma}_{v}$,
we may get \begin{eqnarray}
\eta_{v}^{2}+{\vec{\sigma}_{v}}^{2}=-\frac{1}{2c}\left[-\frac{b^{2}}{2c}\left(p_{0}^{2}+{\vec{s}}^{2}\right)+\left(-\frac{b}{2c}\left(e-f\frac{b+e}{c+f}\right)-e\frac{b+e}{2(c+f)}+2f\left(\frac{b+e}{2(c+f)}\right)^{2}\right)\left(u_{0}p_{0}+\vec{u}\vec{s}\right)^{2}\right]\label{eq:GapRes3}\end{eqnarray}
 Combining results~(\ref{eq:GapRes1}-\ref{eq:GapRes3}), we may
get for the effective action \begin{eqnarray}
 &  & \Gamma_{eff}[\lambda,\sigma,\vec{s},p_{0},u_{i}]=...+a\left(p_{0}^{2}+{\vec{s}}^{2}\right)-\frac{b}{2c}\left[b\left(p_{0}^{2}+{\vec{s}}^{2}\right)+\left(e-f\frac{b+e}{c+f}\right)\left(u_{0}p_{0}+\vec{u}\vec{s}\right)^{2}\right]\label{eq:GammaEffChiChi}\\
 &  & -\left[-\frac{b^{2}}{4c}\left(p_{0}^{2}+{\vec{s}}^{2}\right)+\left(-\frac{b}{4c}\left(e-f\frac{b+e}{c+f}\right)-e\frac{b+e}{4(c+f)}+f\left(\frac{b+e}{2(c+f)}\right)^{2}\right)\left(u_{0}p_{0}+\vec{u}\vec{s}\right)^{2}\right]\nonumber \\
 &  & +d\left(u_{0}p_{0}+\vec{u}\vec{s}\right)^{2}-e\frac{b+e}{2(c+f)}\left(u_{0}p_{0}+\vec{u}\vec{s}\right)^{2}+f\left(\frac{b+e}{2(c+f)}\left(u_{0}p_{0}+\vec{u}\vec{s}\right)\right)^{2}\nonumber \\
 &  & =...+\left(a-\frac{b^{2}}{4c}\right)\left(p_{0}^{2}+{\vec{s}}^{2}\right)+\left(d-\frac{b}{4c}\left(e-f\frac{b+e}{c+f}\right)-e\frac{b+e}{4(c+f)}\right)\left(u_{0}p_{0}+\vec{u}\vec{s}\right)^{2}+\mathcal{O}\left(\chi,\chi^{\dagger}\chi\right),\nonumber \end{eqnarray}
 where we omitted the terms which are proportional to the chiral doublet
$\chi$. The terms shown in~(\ref{eq:GammaEffChiChi}) are explicitly
chiral invariant and correspond to the terms $\left(\tilde{\chi}^{\dagger}\tilde{\chi}\right)$
and $\left(\tilde{\chi}^{\dagger}U\right)^{2}$ in the chiral lagrangian~\cite{Gasser:1983yg}.
Respectively, for the constant $l_{7}$ we may deduce \begin{eqnarray}
l_{7}=\frac{d-\frac{b}{4c}\left(e-f\frac{b+e}{c+f}\right)-e\frac{b+e}{4(c+f)}}{4B^{2}}\label{eq:L7-abcdef}\end{eqnarray}
 Thus we can see that in addition to the term $d$ in numerator there
are three other terms which correspond to contributions of additional
mesons. As we will see from the following sections, these contributions
have different signs and approximately the same order of magnitude
as the term $d$. The formula~(\ref{eq:L7-abcdef}) proves that we
have to consider correlators instead of direct comparison of the terms
in the expansion of the lagrangian.

Below we will not evaluate the constants $\mathcal{A}-\mathcal{D},a-f,$
but instead evaluate the correlators directly.

\section{Gap equation}

\label{sec:GapEqn}

The next-to-leading order (NLO) gap equations which follow from the
effective action~(\ref{eq:Z0}) have a form\begin{eqnarray}
\sigma\frac{\partial S}{\partial\sigma} & = & 4\sigma^{2}-\frac{1}{V}Tr\left(iM(p)\hat{S}(p)\right)-\frac{1}{\sigma^{2}}\int\frac{d^{4}q}{(2\pi)^{4}}\sum V_{3}^{(ij)}(q)\Pi_{ij}(q)=0,\label{eq:GapSigmaNLO}\\
\sigma_{3}\frac{\partial S}{\partial\sigma_{3}} & = & 4\sigma_{3}^{2}-\frac{1}{V}Tr\left(i\delta M(p)\tau_{3}\hat{S}(p)\right)-\frac{1}{\sigma^{2}}\int\frac{d^{4}q}{(2\pi)^{4}}\sum\tilde{V}_{3}^{(ij)}(q)\Pi_{ij}(q)=0\label{eq:GapEquations}\\
\lambda\frac{\partial S}{\partial\lambda} & = & \frac{N}{V}-\frac{1}{2V}Tr\left(\hat{S}(p)\left(iM(p)+i\tau_{3}\delta M(p)\right)\right)+\nonumber \\
 &  & +\frac{1}{2\sigma^{2}}\int\frac{d^{4}q}{(2\pi)^{4}}\sum_{i}\left(V_{2}^{(ij)}(q)-V_{3}^{(ij)}(q)\right)\Pi_{ij}(q)=0,\label{eq:GapLambdaNLO}\end{eqnarray}

where we used notations\begin{eqnarray}
V_{2}^{(gap)(ij)}(q) & = & \frac{1}{\sigma^{2}}\int\frac{d^{4}p}{(2\pi)^{4}}\, Tr\left(\frac{M(p)}{\hat{p}+i\mu(p)+i\tau_{3}\delta\mu(p)}\Gamma_{i}\frac{M(p+q)}{\hat{p}+\hat{q}+i\mu(p+q)+i\tau_{3}\delta\mu(p+q)}\Gamma_{j}\right),\label{eq:V2gap-definition}\\
V_{3}^{(gap)(ij)}(q) & = & \frac{i}{\sigma^{2}}\int\frac{d^{4}p}{(2\pi)^{4}}\, Tr\left(\left(\frac{M(p)}{\hat{p}+i\mu(p)+i\tau_{3}\delta\mu(p)}\right)^{2}\Gamma_{i}\frac{M(p+q)}{\hat{p}+\hat{q}+i\mu(p+q)+i\tau_{3}\delta\mu(p+q)}\Gamma_{j}\right),\label{eq:V3gap-definition}\\
\tilde{V}_{3}^{(gap)(ij)}(q) & = & \frac{i}{\sigma^{2}}\int\frac{d^{4}p}{(2\pi)^{4}}\, Tr\left(\left(\frac{M(p)\delta M(p)\tau_{3}}{\left(\hat{p}+i\mu(p)+i\tau_{3}\delta\mu(p)\right)^{2}}\right)\Gamma_{i}\frac{M(p+q)}{\hat{p}+\hat{q}+i\mu(p+q)+i\tau_{3}\delta\mu(p+q)}\Gamma_{j}\right),\label{eq:V3tildeGap-definition}\end{eqnarray}
 explicit expressions for the vertices~(\ref{eq:V2gap-definition}-\ref{eq:V3tildeGap-definition})
are given in Appendix~\ref{sec:appExplicitExpressions}, and the
propagators used for evaluation are written out in Section~\ref{sec:Propagators}.
In general, equations~(\ref{eq:GapSigmaNLO}-\ref{eq:GapLambdaNLO})
can be solved only numerically.

\subsection{Expansion over $\delta m$ }

\label{sec:Expansion-epsilon} For the special case when $\delta m$
is small, it is possible to solve the equations~(\ref{eq:GapEquations})
making a systematic expansion over the small parameter $\delta m$.
For our purpose it suffices to keep just the first order corrections.
From the first and the third gap equations in~(\ref{eq:GapEquations})
and the structure of the vertices~(\ref{eq:V2gap-definition},\ref{eq:V3gap-definition})
we may conclude that the vacuum expectation values for $\langle\sigma\rangle,\langle\lambda\rangle$
get corrections only in the second order over $\delta m,$ thus in
the first order they remain the same as for $\delta m=0.$ The equation
for the $\langle\sigma_{3}\rangle$ has a form

\begin{align}
\sigma_{3}\frac{\partial S}{\partial\sigma_{3}} & \approx-4\epsilon^{2}\langle\sigma\rangle^{2}-8\epsilon N_{c}\int\frac{d^{4}p}{(2\pi)^{4}}\frac{\left(p^{2}-\mu^{2}(p)\right)M(p)(\delta m+\epsilon M(p))}{\left(p^{2}+\mu^{2}(p)\right)^{2}}-\label{eq:GapSplitApproximate}\\
 & \frac{1}{\sigma^{2}}\int\frac{d^{4}q}{(2\pi)^{4}}\sum\tilde{V}_{3}^{(ij)}(q)\Pi_{ij}(q)=0,\end{align}

where $\epsilon=\frac{i\langle\sigma_{3}\rangle}{\langle\sigma\rangle},$

\begin{eqnarray*}
\tilde{V}_{3}^{(ij)}\Pi_{ij}(q) & \approx & \frac{\epsilon}{2\sigma^{2}}\int\frac{d^{4}p}{(2\pi)^{4}}M^{2}(p)M(p+q)\times\left\{ \right.\\
 & \left(\Pi_{\sigma\sigma}^{(0)}(q)-\Pi_{\sigma_{3}\sigma_{3}}^{(0)}(q)\right) & Tr\left[S_{+}(p)S_{+}(p)S_{+}(p+q)-S_{-}(p)S_{-}(p)S_{-}(p+q)\right]_{\mathcal{O}(\delta m)}+\\
 & 2\Pi_{\sigma\sigma_{3}} & Tr\left[S_{+}(p)S_{+}(p)S_{+}(p+q)+S_{-}(p)S_{-}(p)S_{-}(p+q)\right]_{\delta m=0}-\\
\sum_{i_{\perp}=1,2} & \Pi_{\sigma_{i}}^{(0)}(k) & Tr\left[S_{+}(p)S_{+}(p)S_{-}(p+q)-S_{-}(p)S_{-}(p)S_{+}(p+q)\right]_{\mathcal{O}(\delta m)}-\\
 & \left(\Pi_{\eta\eta}^{(0)}(k)-\Pi_{\phi_{3}\phi_{3}}^{(0)}(k)\right) & Tr\left[S_{+}(p)S_{+}(p)\bar{S}_{+}(p+q)-S_{-}(p)S_{-}(p)\bar{S}_{-}(p+q)\right]_{\mathcal{O}(\delta m)}-\\
 & 2\Pi_{\eta\phi_{3}}(k) & Tr\left[S_{+}(p)S_{+}(p)\bar{S}_{+}(p+q)+S_{-}(p)S_{-}(p)\bar{S}_{-}(p+q)\right]_{\delta m=0}+\\
\sum_{i_{\perp}=1,2} & \Pi_{\phi_{i}}^{(0)}(k) & Tr\left[S_{+}(p)S_{+}(p)\bar{S}_{-}(p+q)-S_{-}(p)S_{-}(p)\bar{S}_{+}(p+q)\right]_{\mathcal{O}(\delta m)}\\
 &  & \left.\right\} ,\end{eqnarray*}

the superscript $(0)$ on the propagators and subscripts on $Tr[...]_{\alpha}$
indicates that the proper propagator is to be taken in the limit $\delta m=0$
or just collecting the first $\mathcal{O}(\delta m)$-correction.
One can notice that (\ref{eq:GapSplitApproximate}) has a form

\[
\epsilon\left(\mathcal{X}\epsilon+\mathcal{Y}\delta m\right)=0,\]

where the coefficients~$\mathcal{X},\mathcal{Y}$ should be evaluated
with account of $\mathcal{O}(1/N_{c})$-corrections. A trivial nonzero
solution is $\epsilon=-\delta m\,\mathcal{Y}/\mathcal{X},$ which
corresponds to \begin{equation}
\delta\mu(p)=\delta m\left(1-M\, f^{2}(p)\mathcal{Y}/\mathcal{X}\right).\label{eq:MassSplitBA}\end{equation}
 While in general case the explicit expression for the formfactor
has a form\begin{equation}
f(p)=-x\frac{d}{dx}\left(I_{0}(x)K_{0}(x)-I_{1}(x)K_{1}(x)\right)_{x=\frac{p\rho}{2}},\label{eq:FF_BesselExact}\end{equation}

in order to speed up the evaluations here and below we consider two
parameterizations of the formfactor. The first one is a simple {}``dipole''
parameterization~(\cite{Diakonov:1985eg}) of the form \begin{equation}
f(p)=\frac{2}{2+p^{2}\rho^{2}},\label{eq:FF_Dipole}\end{equation}
 which coincides with (\ref{eq:FF_BesselExact}) in the region of
small $p\lesssim2/\rho$. The second parameterization has a form \begin{equation}
f(p)=\left.\frac{1}{\sqrt{1+a_{1}x^{2}+a_{2}x^{4}+a_{3}x^{6}}}\right|_{x=\frac{p\rho}{2}},\label{eq:FF_QuasiBessel}\end{equation}
 where the free parameters $a_{1}...a_{3}$ are fitted to~(\ref{eq:FF_BesselExact}),
and (\ref{eq:FF_QuasiBessel}) agrees with (\ref{eq:FF_BesselExact})
both for the small and asymptotically large $p.$ We will refer to
(\ref{eq:FF_Dipole}) and (\ref{eq:FF_QuasiBessel}) as dipole and
quasibessel parameterizations respectively. Direct comparison of the
two close parameterizations is important in order to demonstrate that
the results of this paper are very sensitive to the shape of the instanton.

We summarize results obtained for the constants $\mathcal{X},\mathcal{Y}$
with different parameterizations of formfactor in Table~\ref{tab:AB-Parameters}.
As we can see, the $1/N_{c}$-corrections $\mathcal{X}_{NLO}$~and
$\mathcal{Y}_{NLO}$ are small compared to $\mathcal{X}_{LO}$ and
$\mathcal{Y}_{LO}$ respectively, so the $1/N_{c}$-expansion works
very well here. It is important to note that both in the leading order
and next-to-leading order the mass $\delta\mu(p)$ changes sign for
$p\sim0.5$~GeV, so we have a compensation of the small-$p$ and
large-$p$ regions.

\begin{table}
\begin{tabular}{|c|c|c|c|c|c|c|}
\hline 
 & \textbf{$\mathcal{X}_{LO}\times10^{3}$}  & \textbf{$\mathcal{Y}_{LO}\times10^{2}$}  & \textbf{$-(\mathcal{Y}/\mathcal{X})_{LO}$}  & \textbf{$\mathcal{X}_{NLO}\times10^{3}$}  & \textbf{$\mathcal{Y}_{NLO}\times10^{2}$}  & \textbf{$-(\mathcal{Y}/\mathcal{X})_{LO+NLO}$}\tabularnewline
\hline 
\textbf{Dipole}  & -8.50  & --3.53 & -4.16 & 0.80 & 0.61 & -3.80\tabularnewline
\hline 
\textbf{QuasiBessel}  & -9.04 & -3.33 & -3.68 & 1.02 & 0.69 & -3.30\tabularnewline
\hline
\end{tabular}

\caption{\label{tab:AB-Parameters}Parameters of mass split in different parameterizations
of formfactor. Dipole corresponds to~(\ref{eq:FF_Dipole}). QuasiBessel
corresponds to~(\ref{eq:FF_QuasiBessel}). Dimensions: $[\mathcal{X}_{LO,NLO}]=[GeV^{4}],[\mathcal{Y}_{LO,NLO}]=[GeV^{3}]$.}

\end{table}

\section{Propagators}

\label{sec:Propagators} In this section we would like to discuss
the propagators of the quarks and mesons in presence of the mass split
$\delta m$.

\subsection{Quark propagator}

Here we consider only the leading order quark propagator, the NLO
corrections to the quark propagator will be considered as separate
meson loop corrections to proper correlators. Since the operator $\hat{p}+i\mu(p)+i\delta\mu(p)\tau_{3}$
is diagonal in the flavour space, its inversion is quite straightforward,
with\[
\hat{S}(p)\equiv\frac{1}{\hat{p}+i\mu(p)+i\delta\mu(p)\tau_{3}}=\frac{1-\tau_{3}}{2}\, S_{-}(p)+\frac{1+\tau_{3}}{2}\, S_{+}(p),\]
 \[
S_{\pm}(p)=\frac{1}{\hat{p}+i\mu_{\pm}(p)},\]
 where $\mu_{\pm}(p)=\mu(p)\pm\delta\mu(p)$.

\subsection{Meson propagator}

For evaluations in this paper we have to evaluate the meson propagator
with account of $1/N_{c}$-corrections. However it is important to
note that NLO evaluations are needed only for $\Pi_{\eta\eta}(0)$,$\underset{q^{2}=-m_{\pi}^{2}}{Res}\Pi_{\phi\phi}(q)$,$\Pi_{\eta\phi}(0)$,
all the other components and expressions for $q\not=0$ may be evaluated
in leading order, which significantly simplifies the task. Due to
$\delta m\not=0$ the propagator is nondiagonal in indices $(ij)$--we
get additional transitions $\sigma\leftrightarrow\vec{\sigma}$ and
$\eta\leftrightarrow\vec{\phi}$. Inversion of the propagator is trivial
and gives:

\begin{equation}
\Pi_{00}=\frac{\left(\Pi^{-1}\right)_{33}}{\left(\Pi^{-1}\right)_{00}\left(\Pi^{-1}\right)_{33}-\left(\Pi^{-1}\right)_{03}\left(\Pi^{-1}\right)_{30}},\qquad\Pi_{33}=\frac{\left(\Pi^{-1}\right)_{00}}{\left(\Pi^{-1}\right)_{00}\left(\Pi^{-1}\right)_{33}-\left(\Pi^{-1}\right)_{03}\left(\Pi^{-1}\right)_{30}},\label{eq:MesonInv_00_33}\end{equation}

\begin{equation}
\Pi_{03}=-\frac{\left(\Pi^{-1}\right)_{30}}{\left(\Pi^{-1}\right)_{00}\left(\Pi^{-1}\right)_{33}-\left(\Pi^{-1}\right)_{03}\left(\Pi^{-1}\right)_{30}},\qquad\Pi_{30}=-\frac{\left(\Pi^{-1}\right)_{03}}{\left(\Pi^{-1}\right)_{00}\left(\Pi^{-1}\right)_{33}-\left(\Pi^{-1}\right)_{03}\left(\Pi^{-1}\right)_{30}},\label{eq:MesonInv_03_30}\end{equation}

\[
\Pi_{ij}=\frac{\delta_{ij}}{\left(\Pi^{-1}\right)_{i}},\,\,\,(i,\, j)\not=3\]

where we used a shorthand notation $(0,3)=(\sigma,\,\sigma_{3})$
for positive parity mesons, and $(0,3)=(\eta,\,\phi_{3})$ for negative
parity mesons.

\subsubsection{Leading order}

In the leading order for the components $\left(\Pi^{-1}\right)_{ij}$
we have \begin{equation}
\left(\Pi^{-1}\right)_{ij}=4\delta_{ij}+\frac{1}{\sigma^{2}}Tr\left(\hat{Q}(p)\Gamma_{i}\hat{Q}(p+q)\Gamma_{j}\right)\label{eq:InversePropagator}\end{equation}

where \[
\hat{Q}(p)=iM(p)\hat{S}(p)\equiv\frac{iM(p)}{\hat{p}+i\mu(p)+i\delta\mu(p)\tau_{3}},\]
 and explicit expressions for the components are given in Appendix~\ref{sec:appExplicitExpressions}.

\subsubsection{NLO correction}

As it was discussed earlier, we need a few values for propagators
in the next-to-leading order. Since the NLO evaluations are numerically
slow, from the very beginning we will concentrate on evaluation of
the following quantities:\begin{align*}
\lim_{(q\to0,m\to0,\delta m\to0)}\Pi_{\eta\eta}^{-1}(q),\lim_{(q\to0,m\to0,\delta m\to0)}\frac{\Pi_{\eta\phi}^{-1}(q)}{\delta m},\underset{q^{2}=0}{Res}\Pi_{\phi\phi}(q).\end{align*}
 All the terms which do not contribute to one of these limits will
be omitted. For the sake of brevity, below we use notation\[
Q(p)\equiv\frac{Q_{+}(p)+Q_{-}(p)}{2}\approx\frac{iM(p)}{\hat{p}+i\mu(p)}+\mathcal{O}\left(\delta m^{2}\right)\]

For the pion propagator $\Pi_{\phi\phi}(q)$, we may use the chiral
limit and put $m,\delta m$ to zero. The NLO expression for the pion
propagator has a form

\begin{align*}
\Pi_{\phi\phi}^{(ab)-1}(q) & =\left[4\delta^{ab}+\frac{1}{\sigma^{2}}Tr_{p}\left(Q(p)i\gamma_{5}\tau^{a}Q(p+q)i\gamma_{5}\tau^{b}\right)\right]+\\
 & +\frac{1}{\sigma^{4}}\int\frac{d^{4}k}{\left(2\pi\right)^{4}}\Pi_{ij}(k)\left(2Tr_{p}\left(Q(p)i\gamma_{5}\tau^{a}Q(p+q)\Gamma_{i}Q(p+q+k)\Gamma_{j}Q(p+q)i\gamma_{5}\tau^{b}\right)\right.+\\
 & \left.+Tr_{p}\left(Q(p)i\gamma_{5}\tau^{a}Q(p+q)\Gamma_{i}Q(p+q+k)i\gamma_{5}\tau^{b}Q(p+k)\Gamma_{j}\right)\right)\\
 & -\frac{4}{\sigma^{6}}\int\frac{d^{4}k}{\left(2\pi\right)^{4}}\Pi_{i}(k)\Pi_{j}(k+q)Tr_{p}\left(Q(p)i\gamma_{5}\tau^{a}Q(p+q)\Gamma_{i}Q(p+q+k)\Gamma_{j}\right)\times\\
 & Tr_{p}\left(Q(p)i\gamma_{5}\tau^{b}Q(p-q)\Gamma_{i}Q(p-q-k)\Gamma_{j}\right).\end{align*}
 In complete analogy, for the $\eta$-meson propagator $\Pi_{\eta\eta}(0)$
we have

\begin{align*}
\Pi_{\eta\eta}^{-1}(0) & =\left[4\delta^{ab}+\frac{1}{\sigma^{2}}Tr_{p}\left(Q(p)\gamma_{5}Q(p)\gamma_{5}\right)\right]+\\
 & +\frac{1}{\sigma^{4}}\int\frac{d^{4}k}{\left(2\pi\right)^{4}}\Pi_{ij}(k)\left(2Tr_{p}\left(Q(p)\gamma_{5}Q(p)\Gamma_{i}Q(p+k)\Gamma_{j}Q(p)\gamma_{5}\right)+\right.\\
 & +\left.Tr_{p}\left(Q(p)\gamma_{5}Q(p)\Gamma_{i}Q(p+k)\gamma_{5}Q(p+k)\Gamma_{j}\right)\right)\\
 & -\frac{4}{\sigma^{6}}\int\frac{d^{4}k}{\left(2\pi\right)^{4}}\Pi_{i}(k)\Pi_{j}(k)Tr_{p}\left(Q(p)\gamma_{5}Q(p)\Gamma_{i}Q(p+k)\Gamma_{j}\right)\times\\
 & Tr_{p}\left(Q(p)\gamma_{5}Q(p)\Gamma_{i}Q(p-k)\Gamma_{j}\right),\end{align*}
 and again we can make evaluations in the chiral limit.

The nondiagonal matrix element $\Pi_{\eta\phi}(0)$ is $\mathcal{O}\left(\delta m\right)$,
so we will extract explicitly $\delta m$ and after that the evaluation
of the constant will be done in the chiral limit. Evaluation is quite
tedious since LO propagators have nondiagonal components. The corresponding
expression has a form\[
\Pi_{\eta\phi}^{-1}(0)=\Pi_{\eta\phi}^{(LO)-1}(0)+\Pi_{\eta\phi}^{(1-meson)-1}(0)+\Pi_{\eta\phi}^{(2-meson)-1}(0),\]
 where

\[
\Pi_{\eta\phi}^{(LO)-1}(0)=\left[\frac{1}{2\sigma^{2}}Tr_{p}\left(Q_{+}(p)i\gamma_{5}Q_{+}(p)\gamma_{5}\right)-\frac{1}{2\sigma^{2}}Tr_{p}\left(Q_{-}(p)i\gamma_{5}Q_{-}(p)\gamma_{5}\right)\right],\]

\begin{eqnarray}
\Pi_{\eta\phi}^{(1-mes)-1}(0) & = & \int\frac{d^{4}q}{(2\pi)^{4}}\sum_{ij}\Pi_{ij}(q)V_{ij}^{(1-mes,\eta\phi)}(q)=\int\frac{d^{4}q}{(2\pi)^{4}}\sum_{ij}\Pi_{ij}(q)\frac{i}{2\sigma^{4}}\int\frac{d^{4}p}{(2\pi)^{4}}\times\nonumber \\
 & \times & Tr\left(2Q(p)\Gamma_{\eta}Q(p)\Gamma_{\phi}Q(p)\Gamma_{i}Q(p+q)\Gamma_{j}+Q(p)\Gamma_{\eta}Q(p)\Gamma_{i}Q(p+q)\Gamma_{\phi}Q(p+q)\Gamma_{j}\right),\label{eq:PiEta1Loop}\end{eqnarray}

\begin{equation}
\Pi_{\eta\phi}^{(2-mes)-1}(0)=-\frac{4}{\sigma^{6}}\int\frac{d^{4}q}{\left(2\pi\right)^{4}}\Pi_{ij}(q)\Pi_{kl}(q)V_{ik}^{(\eta)}(q)V_{jl}^{(\phi)}(q),\label{eq:PiEta2Loop-0}\end{equation}
 \begin{align*}
V_{ik}^{(\eta)}(q) & =\left[Tr_{p}\left(Q(p)\gamma_{5}Q(p)\Gamma_{i}Q(p+q)\Gamma_{k}\right)\right],\\
V_{ik}^{(\phi)}(q) & =\left[Tr_{p}\left(Q(p)i\gamma_{5}\tau^{3}Q(p)\Gamma_{i}Q(p+q)\Gamma_{k}\right)\right],\end{align*}
 and explicit expessions for the verices contributing to~(\ref{eq:PiEta1Loop}-\ref{eq:PiEta2Loop-0})
are given in Appendix~\ref{sec:appExplicitExpressions}.

Numerial results of evaluation are presented in Table~\ref{tab:Propagators}.
As we can see, even in the leading order (LO) there is a strong sensitivity
of the propagator $\Pi_{\eta\phi}^{-1}(0)$ to the shape of the instanton
(formfactor $F(p)$). This dependence is discussed in more detail
in Section~\ref{sec:Conclusion}.

\begin{table}
\begin{tabular}{|c|c|c|c|c|c|c|}
\hline 
 & \textbf{LO}  & \textbf{Mass Shift}  & \textbf{Mass Split}  & \textbf{Meson}  & \textbf{All NLO}  & \textbf{Total}\tabularnewline
\hline 
\textbf{QuasiBessel}  &  &  &  &  &  & \tabularnewline
\hline 
$\Pi_{\eta\eta}^{-1}(0)$  & $5.65\times10^{-3}$  & $9.16\times10^{-3}$  & 0  & $-8.88\times10^{-3}$  & $2.86\times10^{-3}$  & $5.93\times10^{-3}$\tabularnewline
\hline 
$-i\Pi_{\eta\phi}^{-1}(0)$  & $-4.73\times10^{-3}$  & $0.88\times10^{-3}$  & $0.19\times10^{-3}$  & $-2.30\times10^{-3}$  & $-1.23\times10^{-3}$  & $-1.89\times10^{-3}$\tabularnewline
\hline 
$F_{\pi}^{2}$  & $1.24\times10^{-2}$  & $-0.59\times10^{-2}$  & 0  & $0.11\times10^{-2}$  & $-0.49\times10^{-2}$  & $0.76\times10^{-2}$\tabularnewline
\hline 
$\langle\bar{q}q\rangle$  & $2.03\times10^{-2}$  & $-0.77\times10^{-2}$  & 0  & $0.31\times10^{-2}$  & $-0.45\times10^{-2}$  & $1.58\times10^{-2}$\tabularnewline
\hline 
$B$  & $1.64$  & ---  & 0  & ---  & ---  & $2.09$\tabularnewline
\hline 
\textbf{Dipole}  &  &  &  &  &  & \tabularnewline
\hline 
$\Pi_{\eta\eta}^{-1}(0)$  & $5.65\times10^{-3}$  & $9.16\times10^{-3}$  & 0  & $-9.72\times10^{-3}$  & $-0.56\times10^{-3}$  & $5.09\times10^{-3}$\tabularnewline
\hline 
$-i\Pi_{\eta\phi}^{-1}(0)$  & $-1.72\times10^{-3}$  & $1.47\times10^{-3}$  & $0.17\times10^{-3}$  & $-4.95\times10^{-3}$  & $-3.32\times10^{-3}$  & $-5.20\times10^{-3}$\tabularnewline
\hline 
$F_{\pi}^{2}$  & $1.36\times10^{-2}$  & $-0.54\times10^{-2}$  & 0  & $0.31\times10^{-2}$  & $-0.23\times10^{-2}$  & $1.12\times10^{-2}$\tabularnewline
\hline 
$\langle\bar{q}q\rangle$  & $2.18\times10^{-2}$  & $-0.72\times10^{-2}$  & 0  & $0.31\times10^{-2}$  & $-0.41\times10^{-2}$  & $1.76\times10^{-2}$\tabularnewline
\hline 
$B$  & $1.60$  & ---  & 0  & ---  & ---  & $1.57$\tabularnewline
\hline
\end{tabular}

\caption{\label{tab:Propagators}In this table we give the numbers obtained
for propagators and other relevant constants. $F_{\pi}^{2}$ is used
for evaluation of $Res\left(\Pi_{\phi}\right),$ $\langle\bar{q}q\rangle$
is used for extraction of constant $B$.}

\end{table}

\section{Quark condensate}

\label{sec:QuarkCondensate}

Due to the mass split $\delta m$ there is a flavour difference for
the quark condensate $\delta\langle\bar{q}q\rangle=\langle\bar{u}u\rangle-\langle\bar{d}d\rangle$
. In the leading order this split is\[
\delta\left\langle \bar{q}q\right\rangle _{LO}=\frac{i}{2}Tr\left(\tau_{3}S(p)\right)=4N_{c}\int\frac{d^{4}p}{\left(2\pi\right)^{4}}\left(\frac{\mu_{+}(p)}{p^{2}+\mu_{+}^{2}(p)}-\frac{\mu_{-}(p)}{p^{2}+\mu_{-}^{2}(p)}\right).\]
 In the NLO evaluation is also quite straightforward, with\begin{align}
\delta\left\langle \bar{q}q\right\rangle _{meson} & =\int\frac{d^{4}q}{\left(2\pi\right)^{4}}\sum_{ij}\Pi_{ij}(q)V_{ij}^{\left(\delta\bar{q}q\right)}(q),\label{eq:qq_NLO-def}\\
V_{ij}^{\left(\delta\bar{q}q\right)}(q) & =-\int\frac{d^{4}p}{\left(2\pi\right)^{4}}M(p)M(p+q)Tr\left(\frac{i\tau_{3}}{2}S(p)\Gamma_{i}S(p+q)\Gamma_{j}S(p)\right).\nonumber \end{align}
 for meson corrections plus corrections from mass shift and mass split
($1/N_{c}$ corrections to $M_{0}$ and $M_{3}$), and explicit expression
for~(\ref{eq:qq_NLO-def}) is given in Appendix~\ref{sec:appExplicitExpressions}.

Results of numerical evaluation are presented in the Table~\ref{tab:qq-split}.
As one can see, due to the large NLO corrections to the mass split
$M_{u}(p)-M_{d}(p)$, the NLO corrections are larger than the LO result.

\begin{table}
\begin{tabular}{|c|c|c|c|c|c|c|}
\hline 
 & \textbf{LO}  & \textbf{Mass Shift}  & \textbf{Mass Split}  & \textbf{Mesons}  & \textbf{All NLO}  & \textbf{LO+NLO}\tabularnewline
\hline 
\textbf{Dipole}  & -0.20  & $6.07\times10^{-2}$  & $1.26\times10^{-2}$  & $1.09\times10^{-3}$  & $7.45\times10^{-2}$  & -0.13\tabularnewline
\hline 
\textbf{QuasiBessel}  & -0.18 & $6.11\times10^{-2}$  & $1.29\times10^{-2}$  & $2.99\times10^{-3}$  & $7.70\times10^{-2}$  & -0.10\tabularnewline
\hline
\end{tabular}

\caption{\label{tab:qq-split}Different contributions to $\frac{\left\langle \bar{u}u\right\rangle -\left\langle \bar{d}d\right\rangle }{m_{u}-m_{d}}$.
LO: Leading order result Mass Split: Contribution due to NLO correction
to mass split $M_{u}(p)-M_{d}(p)$. Mesons: Contribution of mesons,
All NLO: sum of contributions of mesons and mass shift, LO+NLO--final
result.}

\end{table}

Using formula (11.3) from~\cite{Gasser:1983yg}, it is possible to
get for the constant $h_{3}$ an estimate~%
\footnote{Note that the transition form Minkowsky to Euclid requires to change
the signs of all quark condensates, $\langle\bar{q}q\rangle\to-\langle\bar{q}q\rangle$%
}:\[
h_{3}=\frac{\left.\left(\left\langle \bar{u}u\right\rangle -\left\langle \bar{d}d\right\rangle \right)\right|_{\delta m}^{LO+NLO}}{4B^{2}\delta m}=\frac{0.10\,\delta m}{4B^{2}\delta m}\approx5.48\times10^{-3}.\]

\section{Evaluation of the constant $l_{7}$}

\label{sec:P3P0Correlator}

According to \cite{Gasser:1983yg}, it is possible to evaluate the
constant $l_{7}$ from the correlator $\langle P^{3}(x)P^{0}(0)\rangle$
as\begin{equation}
P_{2}(q)=\int d^{4}x\, e^{iqx}\langle P^{3}(x)P^{0}(0)\rangle=\frac{G_{\pi}\tilde{G}_{\pi}}{m_{\pi}^{2}-q^{2}}+\mathcal{O}\left(q^{2}\right)=\frac{8B^{3}\left(m_{u}-m_{d}\right)}{q^{2}-m_{\pi}^{2}}l_{7}+\mathcal{O}\left(m,q^{2}\right),\label{eq:P3P0Definition}\end{equation}
 where the constant $B$ is one of the phenomenological parameters
of the chiral lagrangian (see Table~\ref{tab:Propagators}), the
mass of the pion $m_{\pi}=0$ in the limit $m\to0$ and $m_{u},m_{d}$
are the current quark masses. Since we are interested only in the
residue of the correlator, we should consider only 1-particle reducible
diagrams with pion in the intermediate state.

\begin{figure}
\includegraphics[scale=0.2]{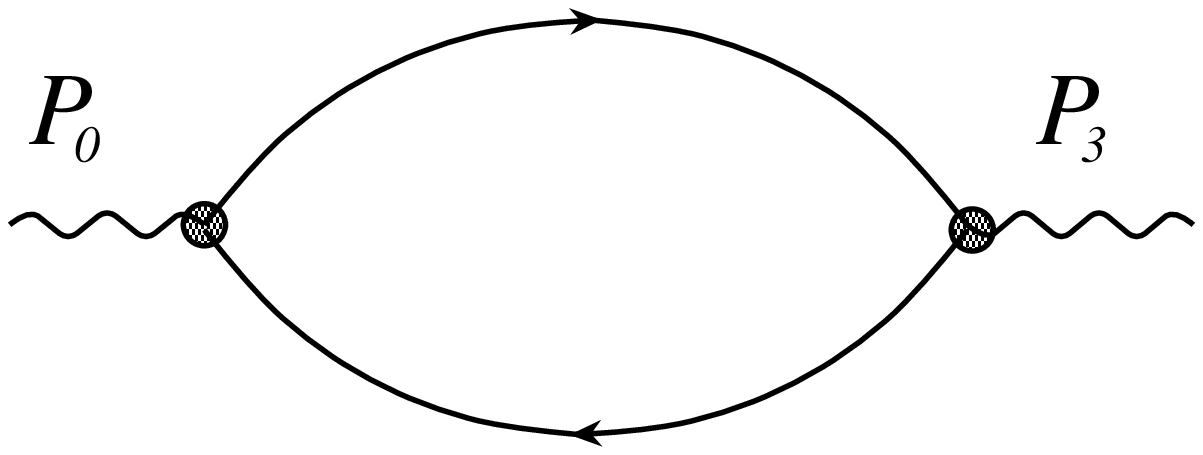}\includegraphics[scale=0.3]{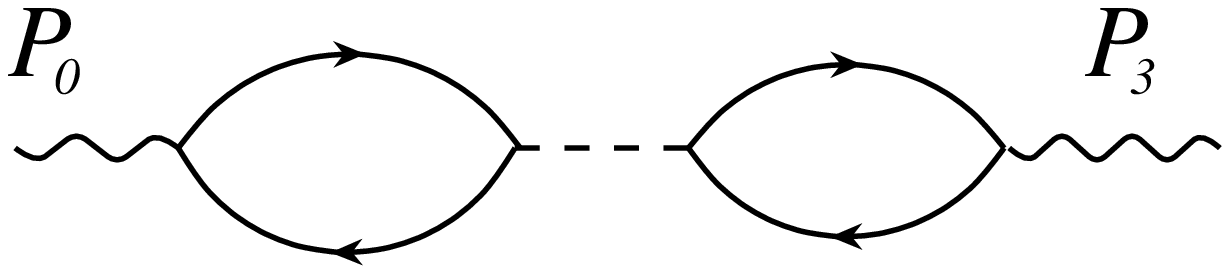}

\caption{\label{fig:P3P0LO}Contribution to the $\langle P^{3}P^{0}\rangle$-correlator
in the leading order.}

\end{figure}

In the leading order, there are two diagrams shown in the Figure~\ref{fig:P3P0LO}.
Obviously, only the diagram on the right-hand side contributes to
the residue, yielding

\[
P_{2}^{LO}(q)=\sum_{i,j=\eta,\phi}L_{i}^{LO}(q)R_{j}^{LO}(q)\Pi_{ij}(q)=L_{\eta}(q)L_{\phi}(q)\left(\Pi_{\eta\eta}(q)+\Pi_{\phi\phi}(q)\right)+(L_{\eta}^{2}(q)+L_{\phi}^{2}(q))\Pi_{\eta\phi}(q),\]
 where

\begin{eqnarray}
L_{\eta}^{LO}(q) & = & -\frac{1}{2}\int\frac{d^{4}p}{(2\pi)^{4}}iMf(p)f(p+q)\left[Tr\left(S_{+}(p)\gamma_{5}S_{+}(p+q)\gamma_{5})\right)+Tr\left(S_{-}(p)\gamma_{5}S_{-}(p+q)\gamma_{5})\right)\right]=\label{eq:P3P0LOLEta}\\
 & = & 4iN_{c}\int\frac{d^{4}p}{(2\pi)^{4}}Mf(p)f(p+q)\left[\frac{p^{2}+p\cdot q+\mu_{+}(p)\mu_{+}(p+q)}{\left(p^{2}+\mu_{+}^{2}(p)\right)\left((p+q)^{2}+\mu_{+}^{2}(p+q)\right)}+\frac{p^{2}+p\cdot q+\mu_{-}(p)\mu_{-}(p+q)}{\left(p^{2}+\mu_{-}^{2}(p)\right)\left((p+q)^{2}+\mu_{-}^{2}(p+q)\right)}\right],\nonumber \\
L_{\phi}^{LO}(q) & = & \frac{1}{2}\int\frac{d^{4}p}{(2\pi)^{4}}Mf(p)f(p+q)\left[Tr\left(S_{+}(p)\gamma_{5}S_{+}(p+q)\gamma_{5})\right)-Tr\left(S_{-}(p)\gamma_{5}S_{-}(p+q)\gamma_{5})\right)\right]=\label{eq:P3P0LOLPhi}\\
 & =- & 4N_{c}\int\frac{d^{4}p}{(2\pi)^{4}}Mf(p)f(p+q)\left[\frac{p^{2}+p\cdot q+\mu_{+}(p)\mu_{+}(p+q)}{\left(p^{2}+\mu_{+}^{2}(p)\right)\left((p+q)^{2}+\mu_{+}^{2}(p+q)\right)}-\frac{p^{2}+p\cdot q+\mu_{-}(p)\mu_{-}(p+q)}{\left(p^{2}+\mu_{-}^{2}(p)\right)\left((p+q)^{2}+\mu_{-}^{2}(p+q)\right)}\right],\nonumber \end{eqnarray}
 and we used identities

\begin{eqnarray}
R_{\eta}^{LO}(q) & = & L_{\phi}^{LO}(q),\label{eq:P3P0LOREta}\\
R_{\phi}^{LO}(q) & = & L_{\eta}^{LO}(q).\label{eq:P3P0LORPhi}\end{eqnarray}

\begin{figure*}
\includegraphics[scale=0.2]{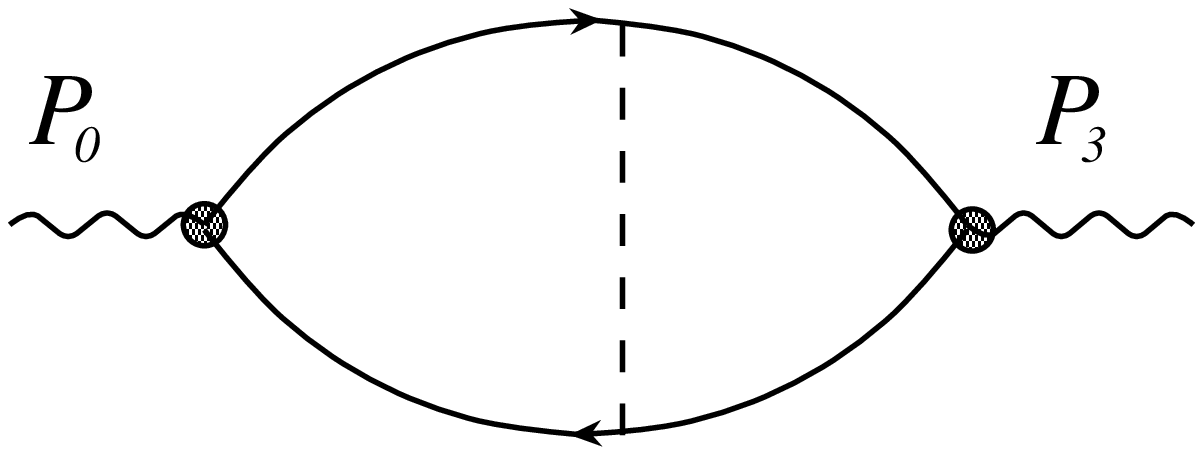}\qquad{}\includegraphics[scale=0.2]{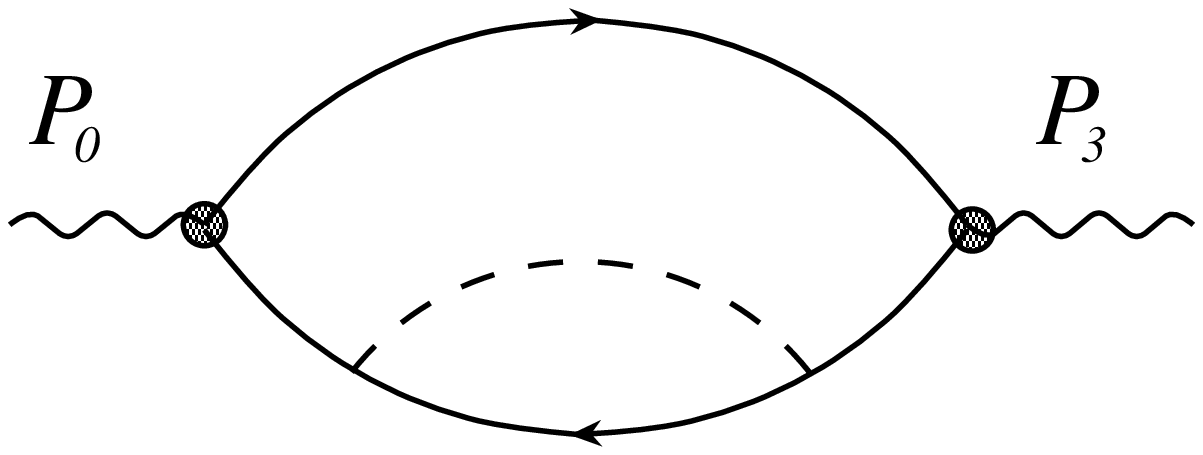}

\includegraphics[scale=0.3]{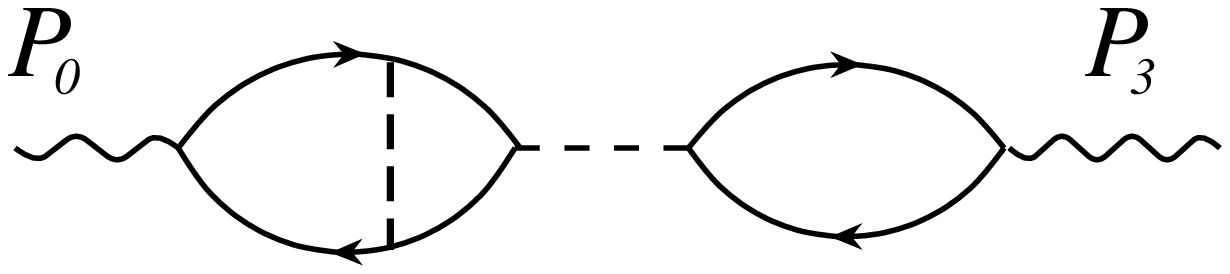}\qquad{}\includegraphics[scale=0.3]{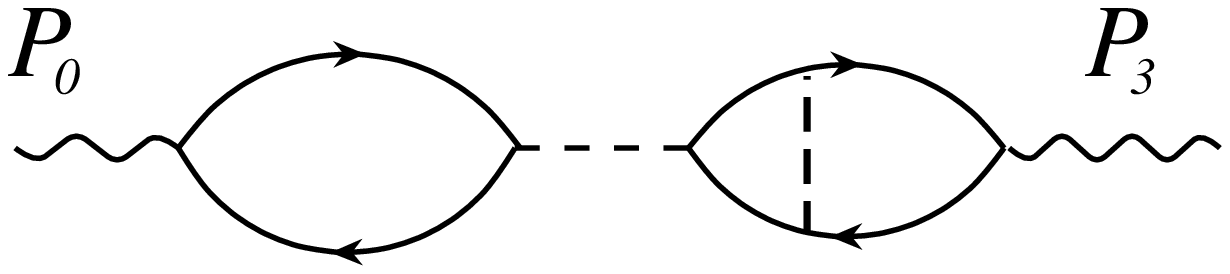}

\includegraphics[scale=0.3]{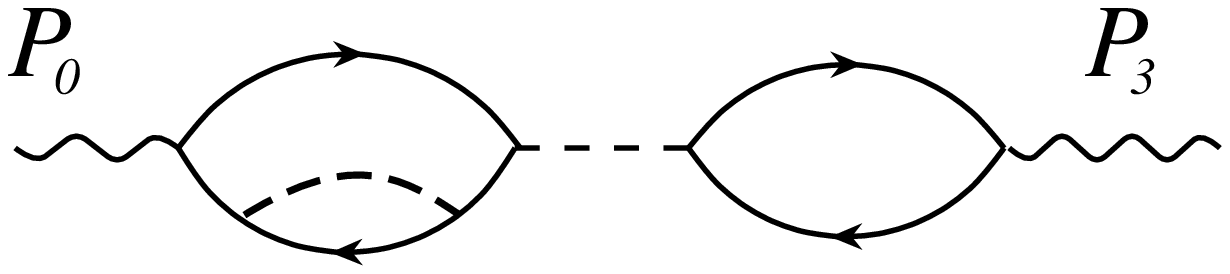}\qquad{}\includegraphics[scale=0.3]{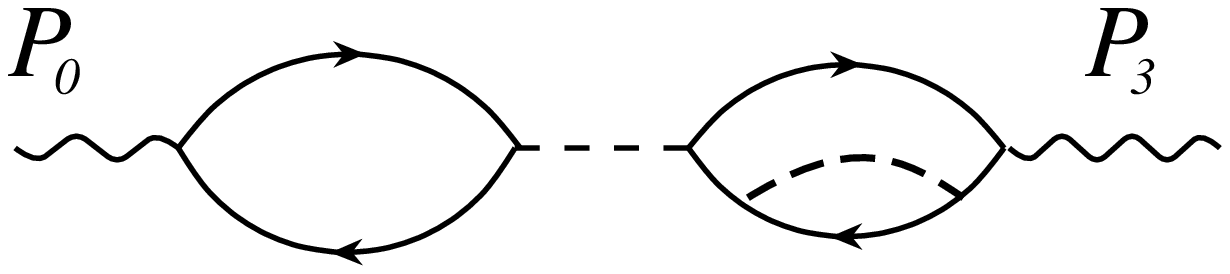}

\qquad{}\includegraphics[scale=0.2]{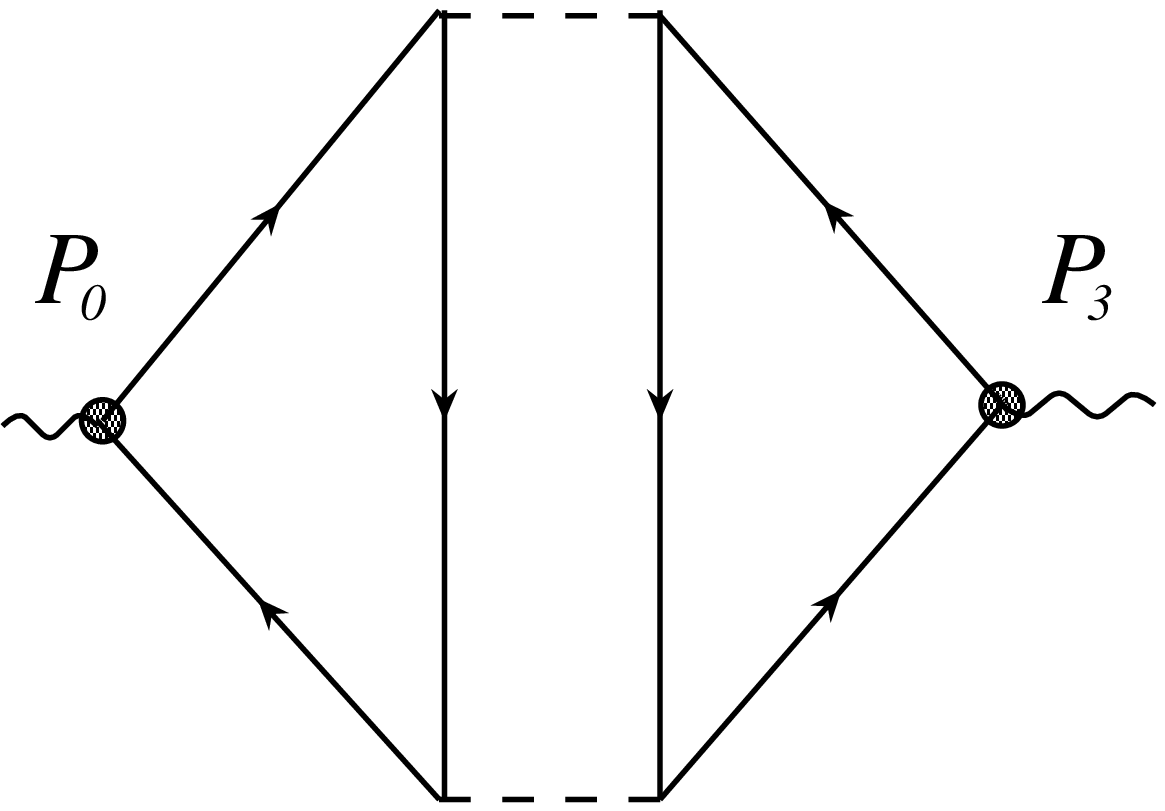}

\caption{\label{fig:P3P0NLO}Contribution to the $\langle P^{3}P^{0}\rangle$-correlator
in the next-to-leading order.}

\end{figure*}

In the next-to-leading order there are seven diagrams shown in the
Figure~\ref{fig:P3P0NLO}. Obviously, only the diagrams 3-6 from
the second and the third row contribute to the residue in pion pole.
The explicit expressions for the corresponding diagrams are given
in Appendix~\ref{sec:appExplicitExpressions}. Using~(\ref{eq:MesonInv_00_33},\ref{eq:MesonInv_03_30}),
one may immediately get

\[
\underset{q^{2}=0}{Res}\Pi_{\eta\phi}(q)\approx-\Pi_{\eta\phi}^{-1}(0)\Pi_{\eta}(0)\underset{q^{2}=0}{Res}\Pi_{\phi\phi}(q),+\mathcal{O}\left(\delta m^{3},m\right)\]

So in evaluation of the residue $\underset{q^{2}=-m_{\pi}^{2}}{Res}\left\langle P_{3}P_{0}\right\rangle \approx\underset{q^{2}=0}{Res}\left\langle P_{3}P_{0}\right\rangle $
one has to keep only the terms

\begin{equation}
\left\langle P_{3}P_{0}\right\rangle =\sum_{i,j=\eta,\phi}L_{i}^{LO}(q)R_{j}^{LO}(q)\Pi_{ij}(q)=L_{\eta}(q)L_{\phi}(q)\Pi_{\phi\phi}(q)+(L_{\eta}^{2}(q)+L_{\phi}^{2}(q))\Pi_{\eta\phi}(q)+non-singulars,\label{eq:ResP3P0Equation}\end{equation}
 all the other terms which are not written out explicitly do not contribute
to the residue.

Results of numerical evaluation are presented in Table~\ref{tab:ResP3P0Values}.
As one can see, the model is extremely sensitive to the change of
formfactor. The reasons of such strong dependence will be discussed
in Section~\ref{sec:Conclusion}.

\begin{table}
\begin{tabular}{|c|c|c|c|c|c|c|}
\hline 
 & \textbf{LO}  & \textbf{Mass shift}  & \textbf{Mass split}  & \textbf{Meson}  & \textbf{All NLO}  & \textbf{Total}\tabularnewline
\hline 
\textbf{QuasiBessel}  &  &  &  &  &  & \tabularnewline
\hline 
$-iL_{\eta}$  & $4.07\times10^{-2}$  & $-1.53\times10^{-2}$  & 0  & $-2.33\times10^{-2}$  & $-3.87\times10^{-2}$  & $1.94\times10^{-3}$\tabularnewline
\hline 
$L_{\phi}$  & $-0.15\times10^{-3}$  & $-8.03\times10^{-3}$  & $-2.34\times10^{-3}$  & $-6.93\times10^{-3}$  & $-1.73\times10^{-2}$  & $-1.74\times10^{-2}$\tabularnewline
\hline 
$l_{7}$  & $0.17\times10^{-4}$  &  &  &  &  & $1.198\times10^{-4}$\tabularnewline
\hline 
\textbf{Dipole}  &  &  &  &  &  & \tabularnewline
\hline 
$-iL_{\eta}$  & $4.35\times10^{-2}$  & $-1.45\times10^{-2}$  & 0  & $-2.38\times10^{-2}$  & $-3.83\times10^{-2}$  & $5.22\times10^{-3}$\tabularnewline
\hline 
$L_{\phi}$  & $6.71\times10^{-3}$  & $-1.06\times10^{-2}$  & $-2.10\times10^{-3}$  & $-1.70\times10^{-2}$  & $-2.97\times10^{-2}$  & $-2.30\times10^{-2}$\tabularnewline
\hline 
$l_{7}$  & $0.34\times10^{-3}$  &  &  &  &  & $1.00\times10^{-3}$\tabularnewline
\hline
\end{tabular}

\caption{\label{tab:ResP3P0Values}Evaluation of the residue $Res_{q^{2}=-m_{\pi}^{2}}\left\langle P_{3}P_{0}\right\rangle $.
See~Eq. (\ref{eq:ResP3P0Equation}) for more details on meaning of
$L_{\eta},L_{\phi}$. The first column is the LO result, columns 2-5
are NLO corrections, column 6 is the total result. In columns 7-8
we give results for $l_{7}$ in LO and NLO (See the Table~\ref{tab:Propagators}
for numbers used in evaluation).}

\end{table}

\section{Conclusion}

\label{sec:Conclusion}

\begin{figure}[h]
\includegraphics[scale=0.4]{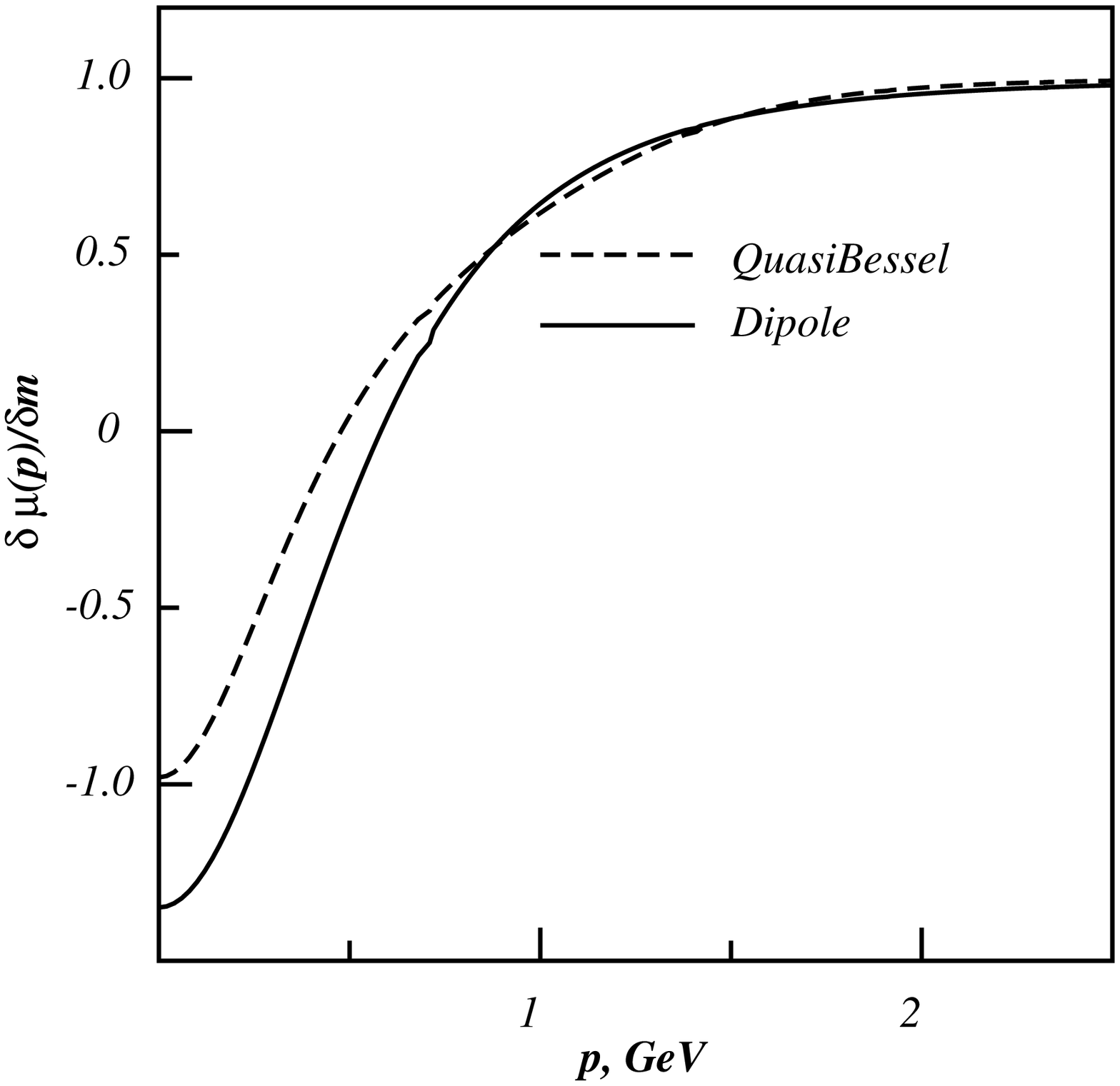}\includegraphics[scale=0.4]{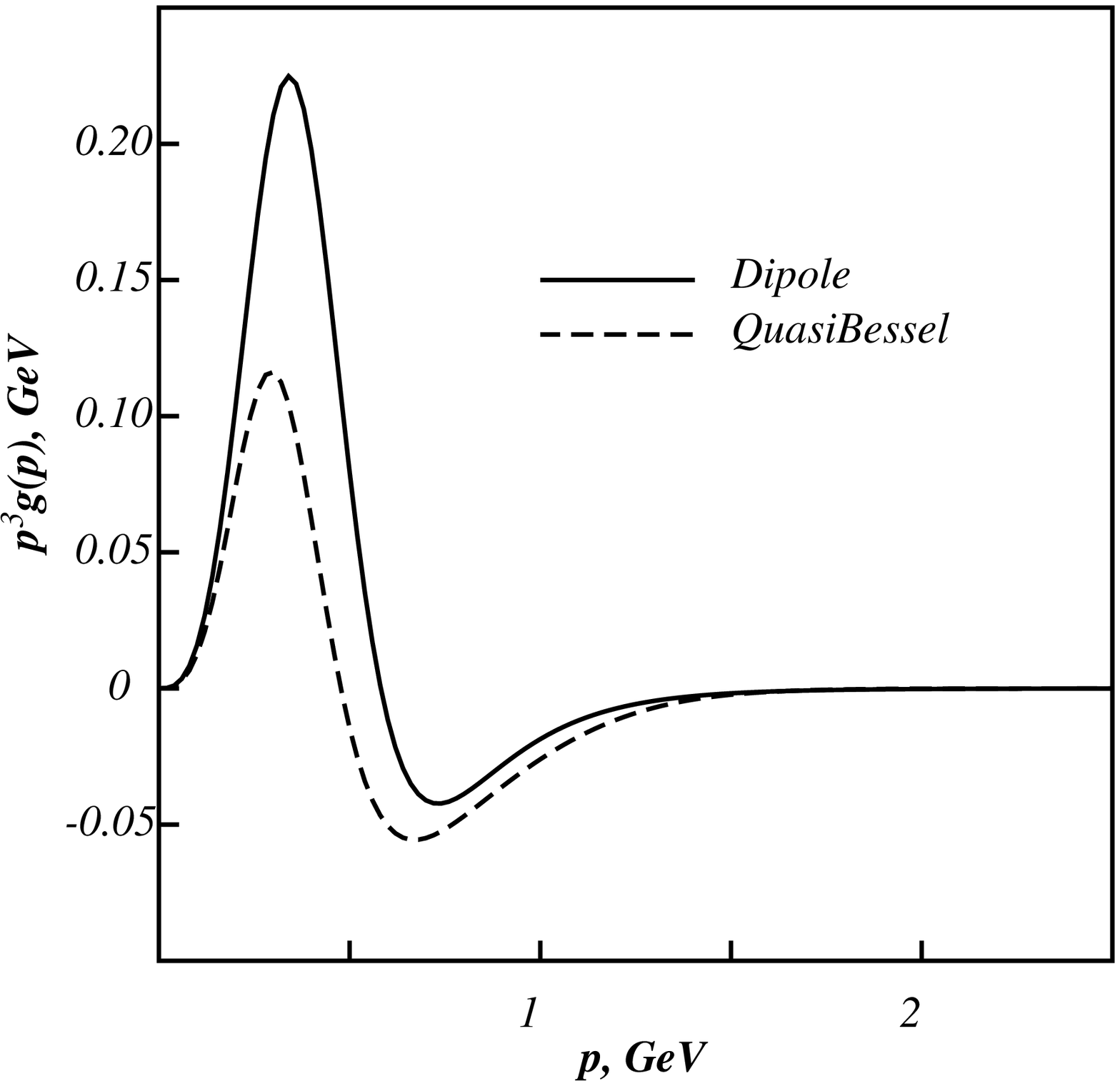}

\caption{\label{fig:ShapeDep}Left: dependence of the dynamical mass split
$\delta\mu(p)\equiv\delta m+\delta Mf^{2}(p)$ on the quark momentum
$p$. Right: Instanton shape dependence of the integrand of $L_{\phi}(0)$
in the leading order. $g(p)$ is the integrand of the Eqn~\ref{eq:P3P0LOLPhi}.}

\end{figure}

In this paper we evaluated the effects of the current quark mass split
on the dynamical mass, quark condensate and correlator $\left\langle P_{3}P_{0}\right\rangle $.
From these data we extracted the low energy constants $h_{3},l_{7}.$
We found that the dynamical quark mass $\delta M$ is negative, so
as one can see from the left pane of the Figure~\ref{fig:ShapeDep},
the momentum-dependent mass $\delta\mu(p)\equiv\delta m+\delta Mf^{2}(p)$
has different signs for small and large momenta. Due to cancellation
of these contributions, we got very strong sensitivity of all quantities
discussed in this paper to the details of the instanton vacuum model,
such as the shape of instanton (which comes via the formfactor) and
instanton parameters. In the right pane of the Figure~\ref{fig:ShapeDep}
we demonstrate explicitly this fast dependence on the example of the
leading-order integrand of $L_{\phi}(0)$. As it was explained above,
due to different signs of large and small-$p$ contributions, we have
partial (solid line) or almost complete (dashed line) cancellation,
which leads to the strong dependence on parameters of the model. Similar
behaviour is observed for all quantities where the dynamical mass
split $\delta M(p)$ contributes, both in the leading and in the next-to-leading
orders.

It is necessary to note that the instanton vacuum model contains chiral
doublet $(\eta,\vec{\sigma})$-additional degree of freedom which
is absent in the chiral lagrangian, and the cancellation of the different
contributions is due to the dynamics of the field. If we set $-\mathcal{Y}/\mathcal{X}=0$
in (\ref{eq:MassSplitBA}) and thus effectively eliminate the contribution
of the $\sigma_{3}$, we can see that the dynamical mass split $\delta\mu(p)$
is constant for all momenta $p$, and cancellation of different regions
does not happen.

\begin{figure}
\includegraphics[scale=0.4,angle=-90]{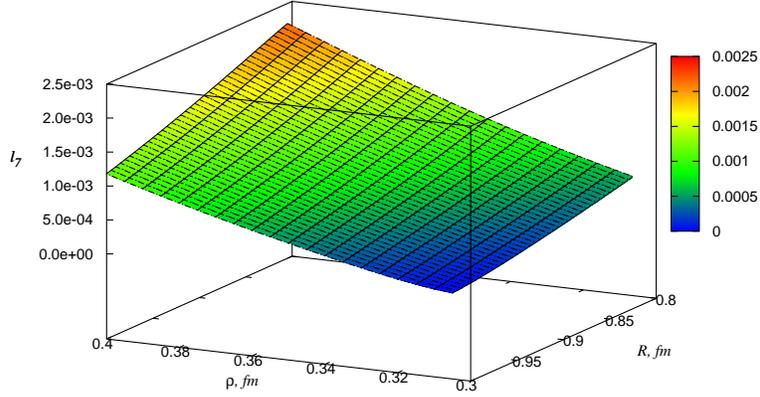}\caption{\label{fig:L7_Rho_R}{[}Color online{]} Dependence of the constant
$l_{7}$ on the instanton vacuum parameters $\rho$ and $R$}

\end{figure}

One of the consequences of the above-mentioned sensitivity of $l_{7}$
to model details is that uncertainty of the instanton vacuum parameters
(average instanton size $\rho$ and inter-instanton distance $R$)
leads to increased uncertainty in the final prediction for $l_{7}$.
As it has been discussed in~\cite{Goeke:2007bj}, different methods
estimate the model parameters are in the range $\rho\sim0.32-0.35$~fm,
$R\sim0.8-1$~fm.While the uncertainty in $\rho,\, R$ is just $\sim10\%$
and is unimportant for most evaluations, for the constant $l_{7}$
it leads to sizeable uncertainty in the final result. Using the Figure~\ref{fig:L7_Rho_R},
we may get for $l_{7}$ an estimate 

\begin{equation}
l_{7}\sim(6.6\pm2.4)\times10^{-4}.\label{eq:L7-final-Value}\end{equation}

The result~(\ref{eq:L7-final-Value}) agrees with a phenomenological
estimate~(\ref{eq:L7-phenom-value}) within uncertainty limits. Using~(\ref{eq:PiMassSplit_QCD}),
we may obtain for the pure QCD contribution to the pion mass difference\[
\left(m_{\pi^{+}}^{2}-m_{\pi^{0}}^{2}\right)_{QCD}\sim1.4\times10^{-5}GeV^{2},\]
i.e. $\sim1\%$ of the experimentally observed difference. This result
does not contradict the well-known fact that the pion mass difference
has electromagnetic origin~\cite{Das:1967it,Langacker:1974nm,Gerstein:1967zz}. 
\begin{acknowledgments}
This work was supported in part by Fondecyt (Chile) grant 1090073,
and by the bilateral Funds DFG-436 USB 113/11/0-1 between Germany
and Uzbekistan. 
\end{acknowledgments}
\appendix

\section{Explicit expressions for some vertices}

\label{sec:appExplicitExpressions}

In this section for the sake of completeness we would like to present
some explicit expressions for the meson-quark interaction vertices
which are used in this paper. For the quark-meson vertices~(\ref{eq:V2gap-definition}-\ref{eq:V3tildeGap-definition})
which come into the gap Eqns.~(\ref{eq:GapSigmaNLO}-\ref{eq:GapLambdaNLO})
we may get

\begin{eqnarray*}
V_{2}^{(gap)(ij)}(q)\Pi_{ij}(q) & = & \frac{1}{\sigma^{2}}\int\frac{d^{4}p}{(2\pi)^{4}}\, Tr\left(\frac{M(p)}{\hat{p}+i\mu(p)+i\tau_{3}\delta\mu(p)}\Gamma_{i}\times\right.\\
 &  & \left.\frac{M(p+q)}{\hat{p}+\hat{q}+i\mu(p+q)+i\tau_{3}\delta\mu(p+q)}\Gamma_{j}\right)=\\
 & = & \frac{1}{2\sigma^{2}}\int\frac{d^{4}p}{(2\pi)^{4}}M(p)M(p+q)\times\left\{ \right.\\
 & \left(\Pi_{\sigma\sigma}(q)-\Pi_{\sigma_{3}\sigma_{3}}(q)\right) & Tr\left[S_{+}(p)S_{+}(p+q)+S_{-}(p)S_{-}(p+q)\right]+\\
 & 2\Pi_{\sigma\sigma_{3}} & Tr\left[S_{+}(p)S_{+}(p+q)-S_{-}(p)S_{-}(p+q)\right]-\\
\sum_{i_{\perp}=1,2} & \Pi_{\sigma_{i}}(k) & Tr\left[S_{+}(p)S_{-}(p+q)+S_{-}(p)S_{+}(p+q)\right]-\\
 & \left(\Pi_{\eta\eta}(k)-\Pi_{\phi_{3}\phi_{3}}(k)\right) & Tr\left[S_{+}(p)\bar{S}_{+}(p+q)+S_{-}(p)\bar{S}_{-}(p+q)\right]-\\
 & 2\Pi_{\eta\phi_{3}}(k) & Tr\left[S_{+}(p)\bar{S}_{+}(p+q)-S_{-}(p)\bar{S}_{-}(p+q)\right]+\\
\sum_{i_{\perp}=1,2} & \Pi_{\phi_{i}}(k) & Tr\left[S_{+}(p)\bar{S}_{-}(p+q)+S_{-}(p)\bar{S}_{+}(p+q)\right]\\
 &  & \left.\right\} \end{eqnarray*}

\begin{eqnarray*}
V_{3}^{(gap)(ij)}\Pi_{ij}(q) & = & \frac{i}{\sigma^{2}}\int\frac{d^{4}p}{(2\pi)^{4}}\, Tr\left(\left(\frac{M(p)}{\hat{p}+i\mu(p)+i\tau_{3}\delta\mu(p)}\right)^{2}\Gamma_{i}\times\right.\\
 &  & \left.\frac{M(p+q)}{\hat{p}+\hat{q}+i\mu(p+q)+i\tau_{3}\delta\mu(p+q)}\Gamma_{j}\right)=\\
 & = & \frac{1}{2\sigma^{2}}\int\frac{d^{4}p}{(2\pi)^{4}}M^{2}(p)M(p+q)\times\left\{ \right.\\
 & \left(\Pi_{\sigma\sigma}(q)-\Pi_{\sigma_{3}\sigma_{3}}(q)\right) & Tr\left[S_{+}(p)S_{+}(p)S_{+}(p+q)+S_{-}(p)S_{-}(p)S_{-}(p+q)\right]+\\
 & 2\Pi_{\sigma\sigma_{3}} & Tr\left[S_{+}(p)S_{+}(p)S_{+}(p+q)-S_{-}(p)S_{-}(p)S_{-}(p+q)\right]-\\
\sum_{i_{\perp}=1,2} & \Pi_{\sigma_{i}}(k) & Tr\left[S_{+}(p)S_{+}(p)S_{-}(p+q)+S_{-}(p)S_{-}(p)S_{+}(p+q)\right]-\\
 & \left(\Pi_{\eta\eta}(k)-\Pi_{\phi_{3}\phi_{3}}(k)\right) & Tr\left[S_{+}(p)S_{+}(p)\bar{S}_{+}(p+q)+S_{-}(p)S_{-}(p)\bar{S}_{-}(p+q)\right]-\\
 & 2\Pi_{\eta\phi_{3}}(k) & Tr\left[S_{+}(p)S_{+}(p)\bar{S}_{+}(p+q)-S_{-}(p)S_{-}(p)\bar{S}_{-}(p+q)\right]+\\
\sum_{i_{\perp}=1,2} & \Pi_{\phi_{i}}(k) & Tr\left[S_{+}(p)S_{+}(p)\bar{S}_{-}(p+q)+S_{-}(p)S_{-}(p)\bar{S}_{+}(p+q)\right]\\
 &  & \left.\right\} \end{eqnarray*}

\begin{eqnarray*}
\tilde{V}_{3}^{(gap)(ij)}\Pi_{ij}(q) & = & \frac{i}{\sigma^{2}}\int\frac{d^{4}p}{(2\pi)^{4}}\, Tr\left(\left(\frac{M(p)\delta M(p)\tau_{3}}{\left(\hat{p}+i\mu(p)+i\tau_{3}\delta\mu(p)\right)^{2}}\right)\Gamma_{i}\times\right.\\
 &  & \left.\frac{M(p+q)}{\hat{p}+\hat{q}+i\mu(p+q)+i\tau_{3}\delta\mu(p+q)}\Gamma_{j}\right)=\\
 & = & \frac{1}{2\sigma^{2}}\int\frac{d^{4}p}{(2\pi)^{4}}M(p)\delta M(p)M(p+q)\times\left\{ \right.\\
 & \left(\Pi_{\sigma\sigma}(q)-\Pi_{\sigma_{3}\sigma_{3}}(q)\right) & Tr\left[S_{+}(p)S_{+}(p)S_{+}(p+q)-S_{-}(p)S_{-}(p)S_{-}(p+q)\right]+\\
 & 2\Pi_{\sigma\sigma_{3}} & Tr\left[S_{+}(p)S_{+}(p)S_{+}(p+q)+S_{-}(p)S_{-}(p)S_{-}(p+q)\right]-\\
\sum_{i_{\perp}=1,2} & \Pi_{\sigma_{i}}(k) & Tr\left[S_{+}(p)S_{+}(p)S_{-}(p+q)-S_{-}(p)S_{-}(p)S_{+}(p+q)\right]-\\
 & \left(\Pi_{\eta\eta}(k)-\Pi_{\phi_{3}\phi_{3}}(k)\right) & Tr\left[S_{+}(p)S_{+}(p)\bar{S}_{+}(p+q)-S_{-}(p)S_{-}(p)\bar{S}_{-}(p+q)\right]-\\
 & 2\Pi_{\eta\phi_{3}}(k) & Tr\left[S_{+}(p)S_{+}(p)\bar{S}_{+}(p+q)+S_{-}(p)S_{-}(p)\bar{S}_{-}(p+q)\right]+\\
\sum_{i_{\perp}=1,2} & \Pi_{\phi_{i}}(k) & Tr\left[S_{+}(p)S_{+}(p)\bar{S}_{-}(p+q)-S_{-}(p)S_{-}(p)\bar{S}_{+}(p+q)\right]\\
 &  & \left.\right\} .\end{eqnarray*}

For the components of the leading order meson propagator~(\ref{eq:InversePropagator})
we may get the following explicit expressions %
\footnote{Notice that off-diagonal component $\Pi_{03}^{-1}$ is real, not imaginary.
This is related to our previous choice of imaginary $\left\langle \sigma_{3}\right\rangle =-i\left|\left\langle \sigma_{3}\right\rangle \right|$ %
} \begin{eqnarray}
\left(\Pi^{-1}\right)_{\sigma\sigma} & = & 4+\frac{1}{2\sigma^{2}}Tr\left(Q_{+}(p)Q_{+}(p+q)\right)+\frac{1}{2\sigma^{2}}Tr\left(Q_{-}(p)Q_{-}(p+q)\right)\label{eq:Propagator:Explicit}\\
\left(\Pi^{-1}\right)_{\sigma_{3}\sigma_{3}} & = & 4-\frac{1}{2\sigma^{2}}Tr\left(Q_{+}(p)Q_{+}(p+q)\right)-\frac{1}{2\sigma^{2}}Tr\left(Q_{-}(p)Q_{-}(p+q)\right)\nonumber \\
\left(\Pi^{-1}\right)_{\sigma_{0}\sigma_{3}} & = & \frac{1}{2\sigma^{2}}Tr\left(Q_{+}(p)Q_{+}(p+q)\right)-\frac{1}{2\sigma^{2}}Tr\left(Q_{-}(p)Q_{-}(p+q)\right)\nonumber \\
\left(\Pi^{-1}\right)_{\eta\eta} & = & 4-\frac{1}{2\sigma^{2}}Tr\left(Q_{+}(p)\tilde{Q}_{+}(p+q)\right)-\frac{1}{2\sigma^{2}}Tr\left(Q_{-}(p)\tilde{Q}_{-}(p+q)\right)\nonumber \\
\left(\Pi^{-1}\right)_{\phi_{3}\phi_{3}} & = & 4+\frac{1}{2\sigma^{2}}Tr\left(Q_{+}(p)\tilde{Q}_{+}(p+q)\right)+\frac{1}{2\sigma^{2}}Tr\left(Q_{-}(p)\tilde{Q}_{-}(p+q)\right)\nonumber \\
\left(\Pi^{-1}\right)_{\eta\phi_{3}} & = & -\frac{1}{2\sigma^{2}}Tr\left(Q_{+}(p)\tilde{Q}_{+}(p+q)\right)+\frac{1}{2\sigma^{2}}Tr\left(Q_{-}(p)\tilde{Q}_{-}(p+q)\right)\nonumber \\
\left(\Pi^{-1}\right)_{ij} & = & 4+\frac{1}{2\sigma^{2}}\sum_{\alpha=\pm}Tr\left(Q_{\alpha}(p)\Gamma_{i}Q_{\alpha}(p+q)\Gamma_{i}\right),\,\,(i,\, j)\not=(0,\,3)\nonumber \end{eqnarray}

where $Q_{\pm}(p)=\frac{iM(p)}{\hat{p}+i\mu(p)\pm i\delta\mu(p)},\,\tilde{Q}_{\pm}(p)\equiv-\gamma_{5}Q_{\pm}(p)\gamma_{5}=\frac{iM(p)}{\hat{p}-i\mu(p)\mp i\delta\mu(p)}.$

The 1-loop correction to the propagator $\Pi_{\eta\phi}^{(1-mes)-1}(0)$
has a form\begin{eqnarray*}
 &  & \Pi_{\eta\phi}^{(1-mes)-1}(0)=\int\frac{d^{4}q}{(2\pi)^{4}}\sum_{ij}\Pi_{ij}(q)V_{ij}(q)=\\
 & = & \int\frac{d^{4}q}{(2\pi)^{4}}\sum_{ij}\Pi_{ij}(q)\frac{i}{2\sigma^{4}}\int\frac{d^{4}p}{(2\pi)^{4}}Tr\left(2Q(p)\Gamma_{\eta}Q(p)\Gamma_{\phi}Q(p)\Gamma_{i}Q(p+q)\Gamma_{j}+Q(p)\Gamma_{\eta}Q(p)\Gamma_{i}Q(p+q)\Gamma_{\phi}Q(p+q)\Gamma_{j}\right)=\\
 &  & \frac{i}{2\sigma^{4}}\int\frac{d^{4}p}{(2\pi)^{4}}M^{2}(p)M(p+q)\times\left\{ \right.\\
 &  & \left(\Pi_{\sigma\sigma}^{(0)}(q)-\Pi_{\sigma_{3}\sigma_{3}}^{(0)}(q)\right)Tr\left(-2M(p)\left(\bar{S}_{+}(p)S_{+}(p)S_{+}(p)S_{+}(p+q)-\bar{S}_{-}(p)S_{-}(p)S_{-}(p)S_{-}(p+q)\right)+\right.\\
 &  & \left.+M(p+q)\left(S_{+}(p)\bar{S}_{+}(p)\bar{S}_{+}(p+q)S_{+}(p+q)-S_{-}(p)\bar{S}_{-}(p)\bar{S}_{-}(p+q)S_{-}(p+q)\right)\right)_{\mathcal{O}(\delta m)}+\\
 &  & 2\Pi_{\sigma\sigma_{3}}(q)Tr\left(-2M(p)\left(\bar{S}_{+}(p)S_{+}(p)S_{+}(p)S_{+}(p+q)+\bar{S}_{-}(p)S_{-}(p)S_{-}(p)S_{-}(p+q)\right)+\right.\\
 &  & \left.+M(p+q)\left(S_{+}(p)\bar{S}_{+}(p)\bar{S}_{+}(p+q)S_{+}(p+q)+S_{-}(p)\bar{S}_{-}(p)\bar{S}_{-}(p+q)S_{-}(p+q)\right)\right)_{\delta m=0}-\\
 &  & \sum_{i_{\perp}=1,2}\Pi_{\sigma_{i}}^{(0)}(q)Tr\left(-2M(p)\left(\bar{S}_{+}(p)S_{+}(p)S_{+}(p)S_{-}(p+q)-\bar{S}_{-}(p)S_{-}(p)S_{-}(p)S_{+}(p+q)\right)+\right.\\
 &  & +\left.M(p+q)\left(S_{+}(p)\bar{S}_{+}(p)\bar{S}_{-}(p+q)S_{-}(p+q)-S_{-}(p)\bar{S}_{-}(p)\bar{S}_{+}(p+q)S_{+}(p+q)\right)\right)_{\mathcal{O}(\delta m)}+\\
 &  & \left(\Pi_{\eta\eta}^{(0)}(q)-\Pi_{\phi_{3}\phi_{3}}^{(0)}(q)\right)Tr\left(2M(p)\left(\bar{S}_{+}(p)S_{+}(p)S_{+}(p)\bar{S}_{+}(p+q)-\bar{S}_{-}(p)S_{-}(p)S_{-}(p)\bar{S}_{-}(p+q)\right)+\right.\\
 &  & \left.M(p+q)\left(S_{+}(p)\bar{S}_{+}(p)S_{+}(p+q)\bar{S}_{+}(p+q)-S_{-}(p)\bar{S}_{-}(p)S_{-}(p+q)\bar{S}_{-}(p+q)\right)\right)_{\mathcal{O}(\delta m)}+\\
 &  & 2\Pi_{\eta\phi_{3}}(q)Tr\left(2M(p)\left(\bar{S}_{+}(p)S_{+}(p)S_{+}(p)\bar{S}_{+}(p+q)+\bar{S}_{-}(p)S_{-}(p)S_{-}(p)\bar{S}_{-}(p+q)\right)+\right.\\
 &  & \left.M(p+q)\left(S_{+}(p)\bar{S}_{+}(p)S_{+}(p+q)\bar{S}_{+}(p+q)+S_{-}(p)\bar{S}_{-}(p)S_{-}(p+q)\bar{S}_{-}(p+q)\right)\right)_{\delta m=0}-\\
 &  & \sum_{i_{\perp}=1,2}\Pi_{\phi_{i}}^{(0)}(q)Tr\left(2M(p)\left(\bar{S}_{+}(p)S_{+}(p)S_{+}(p)\bar{S}_{-}(p+q)-\bar{S}_{-}(p)S_{-}(p)S_{-}(p)\bar{S}_{+}(p+q)\right)+\right.\\
 &  & \left.M(p+q)\left(S_{+}(p)\bar{S}_{+}(p)S_{-}(p+q)\bar{S}_{-}(p+q)-S_{-}(p)\bar{S}_{-}(p)S_{+}(p+q)\bar{S}_{+}(p+q)\right)\right)_{\mathcal{O}(\delta m)}\\
 &  & \left.\right\} ,\end{eqnarray*}
 The two-loop correction to $\Pi_{\eta\phi}(0)$ has a form \begin{equation}
\Pi_{\eta\phi}^{(2-mes)-1}(0)=-\frac{4}{\sigma^{6}}\int\frac{d^{4}q}{\left(2\pi\right)^{4}}\Pi_{ij}(q)\Pi_{kl}(q)V_{ik}^{(\eta)}(q)V_{jl}^{(\phi)}(q),\label{eq:PiEta2Loop}\end{equation}
 where\begin{align*}
V_{ik}^{(\eta)}(q) & =\left[Tr_{p}\left(Q(p)\gamma_{5}Q(p)\Gamma_{i}Q(p+q)\Gamma_{k}\right)\right],\\
V_{ik}^{(\phi)}(q) & =\left[Tr_{p}\left(Q(p)i\gamma_{5}\tau^{3}Q(p)\Gamma_{i}Q(p+q)\Gamma_{k}\right)\right].\end{align*}

In explicit form, with account of $\mathcal{O}\left(\delta m\right)$-counting,
(\ref{eq:PiEta2Loop}) has a form \begin{align}
 & \sum_{ijlk}\Pi_{ij}(q)\Pi_{kl}(q)V_{ik}^{(\eta)}(q)V_{jl}^{(\phi)}(q)=\label{eq:PPSummation-1}\\
 & +V_{\sigma\phi_{3}}^{(\eta)}(q)V_{\sigma\phi_{3}}^{(\phi)}(q)\Pi_{\sigma\sigma}(q)\Pi_{\phi_{3}\phi_{3}}(q)+V_{\eta\sigma_{3}}^{(\eta)}(q)V_{\eta\sigma_{3}}^{(\phi)}(q)\Pi_{\sigma_{3}\sigma_{3}}(q)\Pi_{\eta\eta}(q)\nonumber \\
 & +V_{\sigma\eta}^{(\eta)}(q)V_{\sigma\eta}^{(\phi)}(q)\Pi_{\sigma\sigma}(q)\Pi_{\eta\eta}(q)+V_{\sigma_{3}\phi_{3}}^{(\eta)}(q)V_{\sigma_{3}\phi_{3}}^{(\phi)}(q)\Pi_{\sigma_{3}\sigma_{3}}(q)\Pi_{\phi_{3}\phi_{3}}(q)\\
+ & V_{\sigma_{3}\phi_{3}}^{(\eta)}(q)V_{\sigma\phi_{3}}^{(\phi)}(q)\Pi_{\sigma\sigma_{3}}(q)\Pi_{\phi_{3}\phi_{3}}(q)+V_{\sigma_{3}\phi_{3}}^{(\eta)}(q)V_{\eta\sigma_{3}}^{(\phi)}(q)\Pi_{\sigma_{3}\sigma_{3}}(q)\Pi_{\eta\phi_{3}}(q)\nonumber \\
 & +V_{\sigma\eta}^{(\eta)}(q)V_{\sigma\phi_{3}}^{(\phi)}(q)\Pi_{\sigma\sigma}(q)\Pi_{\eta\phi_{3}}(q)+V_{\sigma\eta}^{(\eta)}(q)V_{\eta\sigma_{3}}^{(\phi)}(q)\Pi_{\sigma\sigma_{3}}(q)\Pi_{\eta\eta}(q)\\
+ & 2V_{\vec{\sigma}_{\perp}\vec{\phi}_{\perp}}^{(\eta)}(q)V_{\vec{\sigma}_{\perp}\vec{\phi}_{\perp}}^{(\phi)}(q)\Pi_{\vec{\sigma}\vec{\sigma}}(q)\Pi_{\vec{\phi}\vec{\phi}}(q)\end{align}
 where the vertices have an explicit form:\begin{align}
V_{\sigma\phi_{3}}^{(\eta)}(q) & =Tr_{p}\left(Q(p)\gamma_{5}Q(p)Q(p+q)i\tau_{3}\gamma_{5}\right)=\nonumber \\
 & -\frac{i}{2}\int\frac{d^{4}p}{\left(2\pi\right)^{4}}M^{2}(p)M\left(p+q\right)Tr_{p}\left(\bar{S}_{+}(p)S_{+}(p)S_{+}(p+q)-\bar{S}_{-}(p)S_{-}(p)S_{-}(p+q)\right)_{\mathcal{O}(\delta m)}\\
V_{\sigma\phi_{3}}^{(\phi)}(q) & =Tr_{p}\left(Q(p)i\gamma_{5}\tau^{3}Q(p)Q(p+q)i\tau_{3}\gamma_{5}\right)=\nonumber \\
 & \frac{1}{2}\int\frac{d^{4}p}{\left(2\pi\right)^{4}}M^{2}(p)M\left(p+q\right)Tr_{p}\left(\bar{S}_{+}(p)S_{+}(p)S_{+}(p+q)+\bar{S}_{-}(p)S_{-}(p)S_{-}(p+q)\right)_{\delta m=0}\\
V_{\eta\sigma_{3}}^{(\eta)}(q) & =Tr_{p}\left(Q(p)\gamma_{5}Q(p)\gamma_{5}Q(p+q)i\tau_{3}\right)=\nonumber \\
 & -\frac{i}{2}\int\frac{d^{4}p}{\left(2\pi\right)^{4}}M^{2}(p)M\left(p+q\right)Tr_{p}\left(S_{+}(p)\bar{S}_{+}(p)S_{+}(p+q)-S_{-}(p)\bar{S}_{-}(p)S_{-}(p+q)\right)_{\mathcal{O}(\delta m)}\\
V_{\eta\sigma_{3}}^{(\phi)}(q) & =Tr_{p}\left(Q(p)i\gamma_{5}\tau^{3}Q(p)\gamma_{5}Q(p+q)i\tau_{3}\right)=\nonumber \\
 & \frac{1}{2}\int\frac{d^{4}p}{\left(2\pi\right)^{4}}M^{2}(p)M\left(p+q\right)Tr_{p}\left(S_{+}(p)\bar{S}_{+}(p)S_{+}(p+q)+S_{-}(p)\bar{S}_{-}(p)S_{-}(p+q)\right)_{\delta m=0}\\
V_{\vec{\sigma}_{\perp}\vec{\phi}_{\perp}}^{(\eta)}(q) & =Tr_{p}\left(Q(p)\gamma_{5}Q(p)i\tau_{\perp}\gamma_{5}Q(p+q)i\tau_{\perp}\gamma_{5}\right)=\\
 & \frac{1}{2}\int\frac{d^{4}p}{\left(2\pi\right)^{4}}M^{2}(p)M\left(p+q\right)Tr_{p}\left(\bar{S}_{+}(p)S_{+}(p)S_{-}(p+q)+\bar{S}_{-}(p)S_{-}(p)S_{+}(p+q)\right)_{\delta m=0}\\
V_{\vec{\sigma}_{\perp}\vec{\phi}_{\perp}}^{(\phi)}(q) & =Tr_{p}\left(Q(p)i\tau_{3}\gamma_{5}Q(p)i\tau_{\perp}\gamma_{5}Q(p+q)i\tau_{\perp}\gamma_{5}\right)=\\
 & \frac{i}{2}\int\frac{d^{4}p}{\left(2\pi\right)^{4}}M^{2}(p)M\left(p+q\right)Tr_{p}\left(\bar{S}_{+}(p)S_{+}(p)S_{-}(p+q)-\bar{S}_{-}(p)S_{-}(p)S_{+}(p+q)\right)_{\mathcal{O}(\delta m)}\\
V_{\sigma\eta}^{(\eta)}(q) & =Tr_{p}\left(Q(p)\gamma_{5}Q(p)Q(p+q)\gamma_{5}\right)=-V_{\sigma\phi_{3}}^{(\phi)}(q)=\mathcal{O}\left(\delta m^{0}\right),\label{eq:VEtaSigmaEta}\\
V_{\sigma\eta}^{(\phi)}(q) & =Tr_{p}\left(Q(p)i\tau_{3}\gamma_{5}Q(p)Q(p+q)\gamma_{5}\right)=V_{\sigma\phi_{3}}^{(\eta)}(q)=\mathcal{O}(\delta m),\label{eq:VPhiSigmaEta}\\
V_{\sigma_{3}\phi_{3}}^{(\eta)}(q) & =Tr_{p}\left(Q(p)\gamma_{5}Q(p)i\tau_{3}Q(p+q)i\tau_{3}\gamma_{5}\right)=V_{\sigma\phi_{3}}^{(\phi)}(q)=\mathcal{O}\left(\delta m^{0}\right),\label{eq:VEtaSigma3Phi3}\\
V_{\sigma_{3}\phi_{3}}^{(\phi)}(q) & =Tr_{p}\left(Q(p)i\tau_{3}\gamma_{5}Q(p)i\tau_{3}Q(p+q)i\tau_{3}\gamma_{5}\right)=-V_{\sigma\phi_{3}}^{(\eta)}(q)=\mathcal{O}\left(\delta m\right).\label{eq:VPhi3Sigma3Phi3}\end{align}

Using the last four equations~(\ref{eq:VEtaSigmaEta}-\ref{eq:VPhi3Sigma3Phi3}),
the two-loop contribution~(\ref{eq:PPSummation-1}) may be cast into
the form\begin{align}
 & \sum_{ijlk}\Pi_{ij}(q)\Pi_{kl}(q)V_{ik}^{(\eta)}(q)V_{jl}^{(\phi)}(q)=\label{eq:PPSummation-2}\\
 & V_{\sigma\phi_{3}}^{(\eta)}(q)V_{\sigma\phi_{3}}^{(\phi)}(q)\Pi_{\sigma\sigma}(q)\left(\Pi_{\phi_{3}\phi_{3}}(q)-\Pi_{\eta\eta}(q)\right)\nonumber \\
+ & V_{\eta\sigma_{3}}^{(\eta)}(q)V_{\eta\sigma_{3}}^{(\phi)}(q)\Pi_{\sigma_{3}\sigma_{3}}(q)\Pi_{\eta\eta}(q)-V_{\sigma\phi_{3}}^{(\eta)}(q)V_{\sigma\phi_{3}}^{(\phi)}(q)\Pi_{\sigma_{3}\sigma_{3}}(q)\Pi_{\phi_{3}\phi_{3}}(q)\nonumber \\
\nonumber \\+ & V_{\sigma\phi_{3}}^{(\phi)}(q)V_{\sigma\phi_{3}}^{(\phi)}(q)\left(\Pi_{\sigma\sigma_{3}}(q)\Pi_{\phi_{3}\phi_{3}}(q)-\Pi_{\sigma\sigma}(q)\Pi_{\eta\phi_{3}}(q)\right)\nonumber \\
+ & V_{\sigma\phi_{3}}^{(\phi)}(q)V_{\eta\sigma_{3}}^{(\phi)}(q)\left(\Pi_{\sigma_{3}\sigma_{3}}(q)\Pi_{\eta\phi_{3}}(q)-\Pi_{\sigma\sigma_{3}}(q)\Pi_{\eta\eta}(q)\right),\nonumber \\
\nonumber \\+ & 2V_{\vec{\sigma}_{\perp}\vec{\phi}_{\perp}}^{(\eta)}(q)V_{\vec{\sigma}_{\perp}\vec{\phi}_{\perp}}^{(\phi)}(q)\Pi_{\vec{\sigma}\vec{\sigma}}(q)\Pi_{\vec{\phi}\vec{\phi}}(q)\nonumber \end{align}
 For the meson loop correction to the quark condensate split~(\ref{eq:qq_NLO-def}),
we have

\begin{eqnarray*}
\sum_{ij}\Pi_{ij}(q)V_{ij}^{(\delta\bar{q}q)}(q) & \approx & \frac{i\epsilon}{2\sigma^{2}}\int\frac{d^{4}p}{(2\pi)^{4}}M(p)M(p+q)\times\left\{ \right.\\
 & \left(\Pi_{\sigma\sigma}^{(0)}(q)-\Pi_{\sigma_{3}\sigma_{3}}^{(0)}(q)\right) & Tr\left[S_{+}(p)S_{+}(p)S_{+}(p+q)-S_{-}(p)S_{-}(p)S_{-}(p+q)\right]_{\mathcal{O}(\delta m)}+\\
 & 2\Pi_{\sigma\sigma_{3}}(q) & Tr\left[S_{+}(p)S_{+}(p)S_{+}(p+q)+S_{-}(p)S_{-}(p)S_{-}(p+q)\right]_{\delta m=0}-\\
\sum_{i_{\perp}=1,2} & \Pi_{\sigma_{i}}^{(0)}(q) & Tr\left[S_{+}(p)S_{+}(p)S_{-}(p+q)-S_{-}(p)S_{-}(p)S_{+}(p+q)\right]_{\mathcal{O}(\delta m)}-\\
 & \left(\Pi_{\eta\eta}^{(0)}(q)-\Pi_{\phi_{3}\phi_{3}}^{(0)}(q)\right) & Tr\left[S_{+}(p)S_{+}(p)\bar{S}_{+}(p+q)-S_{-}(p)S_{-}(p)\bar{S}_{-}(p+q)\right]_{\mathcal{O}(\delta m)}-\\
 & 2\Pi_{\eta\phi_{3}}(q) & Tr\left[S_{+}(p)S_{+}(p)\bar{S}_{+}(p+q)+S_{-}(p)S_{-}(p)\bar{S}_{-}(p+q)\right]_{\delta m=0}+\\
\sum_{i_{\perp}=1,2} & \Pi_{\phi_{i}}^{(0)}(q) & Tr\left[S_{+}(p)S_{+}(p)\bar{S}_{-}(p+q)-S_{-}(p)S_{-}(p)\bar{S}_{+}(p+q)\right]_{\mathcal{O}(\delta m)}\\
 &  & \left.\right\} \end{eqnarray*}
 In complete analogy we may evaluate the diagrams shown in the Figure~(\ref{fig:P3P0NLO})
and get the following explicit vertices

\subsubsection{Diagram \#3}

Left part:\begin{eqnarray*}
L_{\eta}^{(3)} & = & 4\, N_{c}\int\frac{d^{4}p}{(2\pi)^{4}}M^{3}f(p)f(p+q)f^{2}(p+k)f^{2}(p+q+k)\times\left\{ \right.\\
 & \left(\Pi_{\sigma\sigma}(k)-\Pi_{\sigma_{3}\sigma_{3}}(k)\right) & Tr\left[\bar{S}_{+}(p)\bar{S}_{+}(p+k)S_{+}(p+q+k)S_{+}(p+q)+\bar{S}_{-}(p)\bar{S}_{-}(p+k)S_{-}(p+q+k)S_{-}(p+q)\right]+\\
 & 2\Pi_{\sigma\sigma_{3}}(k) & Tr\left[\bar{S}_{+}(p)\bar{S}_{+}(p+k)S_{+}(p+q+k)S_{+}(p+q)-\bar{S}_{-}(p)\bar{S}_{-}(p+k)S_{-}(p+q+k)S_{-}(p+q)\right]-\\
\sum_{i_{\perp}=1,2} & \Pi_{\sigma_{i}}(k) & Tr\left[\bar{S}_{+}(p)\bar{S}_{-}(p+k)S_{-}(p+q+k)S_{+}(p+q)+\bar{S}_{-}(p)\bar{S}_{+}(p+k)S_{+}(p+q+k)S_{-}(p+q)\right]+\\
 & \left(\Pi_{\eta\eta}(k)-\Pi_{\phi_{3}\phi_{3}}(k)\right) & Tr\left[S_{+}(p)\bar{S}_{+}(p+k)S_{+}(p+q+k)\bar{S}_{+}(p+q)+S_{-}(p)\bar{S}_{-}(p+k)S_{-}(p+q+k)\bar{S}_{-}(p+q)\right]+\\
 & 2\Pi_{\eta\phi_{3}}(k) & Tr\left[S_{+}(p)\bar{S}_{+}(p+k)S_{+}(p+q+k)\bar{S}_{+}(p+q)-S_{-}(p)\bar{S}_{-}(p+k)S_{-}(p+q+k)\bar{S}_{-}(p+q)\right]-\\
\sum_{i_{\perp}=1,2} & \Pi_{\phi_{i}}(k) & Tr\left[S_{+}(p)\bar{S}_{-}(p+k)S_{-}(p+q+k)\bar{S}_{+}(p+q)+S_{-}(p)\bar{S}_{+}(p+k)S_{+}(p+q+k)\bar{S}_{-}(p+q)\right]\\
 &  & \left.\right\} \end{eqnarray*}

\begin{eqnarray*}
L_{\phi}^{(3)} & = & 4\, N_{c}\int\frac{d^{4}p}{(2\pi)^{4}}M^{3}f(p)f(p+q)f^{2}(p+k)f^{2}(p+q+k)\times\left\{ \right.\\
 & \left(\Pi_{\sigma\sigma}(k)-\Pi_{\sigma_{3}\sigma_{3}}(k)\right) & Tr\left[\bar{S}_{+}(p)\bar{S}_{+}(p+k)S_{+}(p+q+k)S_{+}(p+q)-\bar{S}_{-}(p)\bar{S}_{-}(p+k)S_{-}(p+q+k)S_{-}(p+q)\right]+\\
 & 2\Pi_{\sigma\sigma_{3}}(k) & Tr\left[\bar{S}_{+}(p)\bar{S}_{+}(p+k)S_{+}(p+q+k)S_{+}(p+q)+\bar{S}_{-}(p)\bar{S}_{-}(p+k)S_{-}(p+q+k)S_{-}(p+q)\right]+\\
\sum_{i_{\perp}=1,2} & \Pi_{\sigma_{i}}(k) & Tr\left[\bar{S}_{+}(p)\bar{S}_{-}(p+k)S_{-}(p+q+k)S_{+}(p+q)-\bar{S}_{-}(p)\bar{S}_{+}(p+k)S_{+}(p+q+k)S_{-}(p+q)\right]+\\
 & \left(\Pi_{\eta\eta}(k)-\Pi_{\phi_{3}\phi_{3}}(k)\right) & Tr\left[S_{+}(p)\bar{S}_{+}(p+k)S_{+}(p+q+k)\bar{S}_{+}(p+q)-S_{-}(p)\bar{S}_{-}(p+k)S_{-}(p+q+k)\bar{S}_{-}(p+q)\right]+\\
 & 2\Pi_{\eta\phi_{3}}(k) & Tr\left[S_{+}(p)\bar{S}_{+}(p+k)S_{+}(p+q+k)\bar{S}_{+}(p+q)+S_{-}(p)\bar{S}_{-}(p+k)S_{-}(p+q+k)\bar{S}_{-}(p+q)\right]+\\
\sum_{i_{\perp}=1,2} & \Pi_{\phi_{i}}(k) & Tr\left[S_{+}(p)\bar{S}_{-}(p+k)S_{-}(p+q+k)\bar{S}_{+}(p+q)-S_{-}(p)\bar{S}_{+}(p+k)S_{+}(p+q+k)\bar{S}_{-}(p+q)\right]\\
 &  & \left.\right\} \end{eqnarray*}

Right part has been evaluated in~(\ref{eq:P3P0LOREta},\ref{eq:P3P0LORPhi}).

\subsubsection{Diagram \#4}

Left part has been evaluated in (\ref{eq:P3P0LOLEta},\ref{eq:P3P0LOLPhi}).

Right part:

\begin{eqnarray*}
R_{\eta}^{(4)}(q) & = & L_{\phi}^{(3)}(q)\\
R_{\phi}^{(4)}(q) & = & L_{\eta}^{(3)}(q)\end{eqnarray*}

\subsubsection{Diagram \#5}

Left part:\begin{eqnarray*}
L_{\eta}^{(5)}(q) & = & -8\, N_{c}\int\frac{d^{4}p}{(2\pi)^{4}}M^{3}f^{3}(p)f^{2}(p+k)f(p+q)\times\left\{ \right.\\
 & \left(\Pi_{\sigma\sigma}(k)-\Pi_{\sigma_{3}\sigma_{3}}(k)\right) & Tr\left[S_{+}(p)S_{+}(p+k)S_{+}(p)\bar{S}_{+}(p+q)+S_{-}(p)S_{-}(p+k)S_{-}(p)\bar{S}_{-}(p+q)\right]+\\
 & 2\Pi_{\sigma\sigma_{3}}(k) & Tr\left[S_{+}(p)S_{+}(p+k)S_{+}(p)\bar{S}_{+}(p+q)-S_{-}(p)S_{-}(p+k)S_{-}(p)\bar{S}_{-}(p+q)\right]-\\
\sum_{i_{\perp}=1,2} & \Pi_{\sigma_{i}}(k) & Tr\left[S_{+}(p)S_{-}(p+k)S_{+}(p)\bar{S}_{+}(p+q)+S_{-}(p)S_{+}(p+k)S_{-}(p)\bar{S}_{-}(p+q)\right]-\\
 & \left(\Pi_{\eta\eta}(k)-\Pi_{\phi_{3}\phi_{3}}(k)\right) & Tr\left[S_{+}(p)\bar{S}_{+}(p+k)S_{+}(p)\bar{S}_{+}(p+q)+S_{-}(p)\bar{S}_{-}(p+k)S_{-}(p)\bar{S}_{-}(p+q)\right]-\\
 & 2\Pi_{\eta\phi_{3}}(k) & Tr\left[S_{+}(p)\bar{S}_{+}(p+k)S_{+}(p)\bar{S}_{+}(p+q)-S_{-}(p)\bar{S}_{-}(p+k)S_{-}(p)\bar{S}_{-}(p+q)\right]+\\
\sum_{i_{\perp}=1,2} & \Pi_{\phi_{i}}(k) & Tr\left[S_{+}(p)\bar{S}_{-}(p+k)S_{+}(p)\bar{S}_{+}(p+q)+S_{-}(p)\bar{S}_{+}(p+k)S_{-}(p)\bar{S}_{-}(p+q)\right]\\
 &  & \left.\right\} \end{eqnarray*}

\begin{eqnarray*}
L_{\phi}^{(5)}(q) & = & -8\, N_{c}\int\frac{d^{4}p}{(2\pi)^{4}}M^{3}f^{3}(p)f^{2}(p+k)f(p+q)\times\left\{ \right.\\
 & \left(\Pi_{\sigma\sigma}(k)-\Pi_{\sigma_{3}\sigma_{3}}(k)\right) & Tr\left[S_{+}(p)S_{+}(p+k)S_{+}(p)\bar{S}_{+}(p+q)-S_{-}(p)S_{-}(p+k)S_{-}(p)\bar{S}_{-}(p+q)\right]+\\
 & 2\Pi_{\sigma\sigma_{3}}(k) & Tr\left[S_{+}(p)S_{+}(p+k)S_{+}(p)\bar{S}_{+}(p+q)+S_{-}(p)S_{-}(p+k)S_{-}(p)\bar{S}_{-}(p+q)\right]-\\
\sum_{i_{\perp}=1,2} & \Pi_{\sigma_{i}}(k) & Tr\left[S_{+}(p)S_{-}(p+k)S_{+}(p)\bar{S}_{+}(p+q)-S_{-}(p)S_{+}(p+k)S_{-}(p)\bar{S}_{-}(p+q)\right]-\\
 & \left(\Pi_{\eta\eta}(k)-\Pi_{\phi_{3}\phi_{3}}(k)\right) & Tr\left[S_{+}(p)\bar{S}_{+}(p+k)S_{+}(p)\bar{S}_{+}(p+q)-S_{-}(p)\bar{S}_{-}(p+k)S_{-}(p)\bar{S}_{-}(p+q)\right]-\\
 & 2\Pi_{\eta\phi_{3}}(k) & Tr\left[S_{+}(p)\bar{S}_{+}(p+k)S_{+}(p)\bar{S}_{+}(p+q)+S_{-}(p)\bar{S}_{-}(p+k)S_{-}(p)\bar{S}_{-}(p+q)\right]+\\
\sum_{i_{\perp}=1,2} & \Pi_{\phi_{i}}(k) & Tr\left[S_{+}(p)\bar{S}_{-}(p+k)S_{+}(p)\bar{S}_{+}(p+q)-S_{-}(p)\bar{S}_{+}(p+k)S_{-}(p)\bar{S}_{-}(p+q)\right]\\
 &  & \left.\right\} \end{eqnarray*}

Right part has been evaluated in~(\ref{eq:P3P0LOREta},\ref{eq:P3P0LORPhi}).

\subsubsection{Diagram \#6}

Left part has been evaluated in (\ref{eq:P3P0LOLEta},\ref{eq:P3P0LOLPhi}).

Right part: \begin{eqnarray*}
R_{\eta}^{(6)}(q) & = & L_{\phi}^{(5)}(q)\\
R_{\phi}^{(6)}(q) & = & L_{\eta}^{(5)}(q)\end{eqnarray*}

\end{document}